\documentclass[aps,prl,showpacs,twocolumn,superscriptaddress]{revtex4}
\usepackage{graphicx}
\usepackage{bm}
\usepackage{amsmath}
\usepackage{amsfonts}

\begin{document}
\title{Spinon Phonon Interaction and Ultrasonic Attenuation in Quantum Spin Liquids}

\author{Yi Zhou}
\affiliation{Department of Physics and Zhejiang Institute of Modern Physics, Zhejiang University, Hangzhou, 310027, P.R. China}
%\email{yizhou@zju.edu.cn}

\author{Patrick A Lee}
%\email{}
\affiliation{Department of Physics, Massachusetts Institute of Technology,
77 Massachusetts Avenue, Cambridge, MA 02139}

\begin{abstract}
Several experimental candidates for quantum spin liquids have been discovered in the past few years which appear to support gapless fermionic $S = {1\over 2}$ excitations called spinons.  The spinons may form a Fermi sea coupled to a $U(1)$ gauge field, and may undergo a pairing instability.  We show that despite being charge neutral, the spinons couple to phonons in exactly the same way that electrons do in the long wavelength limit.  Therefore we can use sound attenuation to measure the spinon mass and lifetime.  Furthermore, transverse ultrasonic attenuation is a direct probe of the onset of pairing because the Meissner effect of the gauge field causes a ``rapid fall'' of the attenuation at $T_c$ in addition to the reduction due to the opening of the energy gap.  This phenomenon, well known in clean superconductors, may reveal the existence of the $U(1)$ gauge field.
\end{abstract}

\maketitle
Quantum spin liquid in dimensions greater than one is a long sought state of matter which has eluded experimental 
investigation until recently.\cite{Lee08} We define the quantum spin liquid as an insulator with an odd number of electrons 
per unit cell which does not order magnetically down to zero temperature due to quantum fluctuations.  
The theory of quantum spin liquid is rather well developed, and it is expected that if such a state exists, 
it will have various exotic properties.  For example, the low energy excitations may be objects which may carry 
spin ${1\over 2}$ and no charge, called spinons.  The spinons may be gapped or gapless and may obey either boson 
or fermion statistics.  They will be accompanied by gauge fields, which may be of the $U(1)$ or $Z_2$ variety.  
In the past few years, several candidates for the quantum spin liquid have emerged.  The best studied example 
is a family of organic compounds.  The original $\kappa - $(ET)$_2$Cu$_2$(CN)$_3$ salt (abbreviated as ET) \cite{Kanoda03} has recently been joined by a second material,\cite{Itou08} the Pd(dmit)$_2$(EtMe$_3$Sb) which we shall refer to as dmit.
%Here (ET)$_2$ and Pd(dmit)$_2$ are organic molecules which form dimers. The dimers are arranged in approximately 
%triangular layers separated by non-magnetic charge donor layers made up of molecules of Cu$_2$(CN)$_3$ and (EtMe$_3$Sb), 
%respectively (Et = C$_2$H$_5$, Me = CH$_3$).  From here on we %shall refer to these systems as ET and dmit salts, 
%respectively.  
Both materials are Mott insulators on an approximate triangular lattice with spin $\frac{1}{2}$ per unit cell, but are not far from the 
Mott transition because they become superconductor (ET) or metal (dmit) under modest pressure.  There is no sign of 
magnetic ordering down to 30~mK despite an exchange interaction $J \approx 250$~K.  
Both materials show a linear T coefficient 
of the specific heat at low temperatures, and a finite spin susceptibility.\cite{SYamashita}  The Wilson ratio is close to one, 
usually associated with metals and is highly unusual for an insulator.  The thermal conductivity $\kappa$ is a good probe 
of these low lying excitations.  Experiments on the ET salts indeed found a large contribution, but $\kappa/T$ is reduced 
below 0.3~K.\cite{MYamashita09}  On the other hand, recent experiments on dmit found that $\kappa/T$ extrapolates to a constant 
down to the lowest temperature.\cite{MYamashita10}  These data strongly support the picture that the low lying excitations are 
mobile fermionic particles, called spinons.

Initial theoretical work pointed to a state where spinons form a Fermi surface and are coupled to $U(1)$ gauge fields.\cite{Motrunich,SSLee}  
Due to the proximity to the Mott transition, ring exchange or charge fluctuations yield additional terms to 
the Heisenberg model and help stabilize the spin liquid state.  However, a peak in the specific heat around 6~K in ET 
and 4~K in dmit suggests a phase transition, which in the case of ET, has been confirmed by thermal expansion measurements.\cite{Manna}  
Furthermore, the nuclear spin relaxation rate $1/T_1T$ shows a power law decrease below 1~K.\cite{Itou10}  These data led to 
the suggestion that the Fermi surface may be unstable to a pairing instability which nevertheless leaves a finite 
density of states intrinsically or due to impurities.\cite{Grover}  Thus the true ground state in the organic salts remains unknown at present.

Two other examples, the Kagome compound ZnCu$_3$(OH)$_6$Cl$_2$ and the three dimensional hyper-Kagome Na$_4$Ir$_3$O$_8$ 
also satisfy the condition of being spin liquids in that they do not show magnetic order and both are characterized 
by gapless excitations.\cite{Kagome07,Takagi07} However, less detailed data are available and we shall focus our attention on the organics, 
even though the conceptual question we raise below will apply equally to these materials if fermionic spinons are found to be present.  Spinons with Dirac sectra, however, may require a different treatment.

In this paper we address two questions.  First, how do the spinons couple to phonons, and, secondly, is there a way 
to unambiguously identify the pairing transition of spinons?  As we shall see, the two questions are related because 
the attenuation of transverse  sound turns out to be sensitive to the gauge magnetic field fluctuations and is 
a sensitive probe of the Meissner effect of gauge magnetic field at the onset of any pairing instability.

The coupling of electrons to phonons is often discussed in terms of the screened Coulomb coupling between electrons 
and nuclei and one may have the impression that the charge neutral spinon may couple differently.  It turns out that 
in the long wavelength limit, the coupling matrix elements are exactly the same.  This is because in this limit 
the coupling can be viewed as a distortion of the Fermi surface by the local stress of the unit cell and is the same 
whether the fermions are charged or not.  Recently the spin phonon coupling was discused 
%in the short wavelength limit, 
in terms of interactions mediated by gauge fields 
with the conclusion that the coupling is 
%also 
comparable to the electron phonon coupling.\cite{Mross}  
%Thus we can conclude that 
%the spinon phonon coupling is similar to electron phonon at all length scales.  
The present approach makes it clear that for a spinon Fermi surface, the result does not rely on gauge fields in the long wavelength limit.  
We next consider the relaxation of 
ultrasound due to spinons in parallel with the standard treatment of electron relaxation, including the onset of 
superconductivity.  Differences and similarities will be pointed out.

We begin with a derivation of the spinon phonon coupling following Blount's discussion of the electron phonon problem 
which was also used in Tsuneto's theory of ultrasound attenuation.\cite{Blount,Tsuneto}  Blount's key insight is to transform to 
a frame moving with the lattice distortion.  In a slight departure from Blount, we assume that the kinetic energy of 
the spinon is described by a mean field band $E_0(\bm k)$.  $E_0(\bm k)$ can be a nearest neighbor tight binding band, 
for instance.  The unperturbed Hamiltonian is
\begin{equation}
H_0 = E_0(\bm p) + V_{\rm imp}({\bm r}^\prime)  ,
\end{equation}
where ${\bm p} = -i{\partial \over \partial{\bm r}}$ is the momentum operator, $V_{\rm imp}$ describes disorder scattering 
which relaxes the spinon distribution to the lattice, and ${\bm r}^\prime$ refers to the laboratory frame.  
The sound wave is described by $\delta {\bm R}({\bm r}^\prime,t)$ which is a smoothly varying function of ${\bm r}^\prime$ 
such that it equals the displacement of the ions at the lattice points.  The transformation to the moving frame 
${\bm r} = {\bm r}^\prime - \delta{\bm R}({\bm r^{\prime}},t)$ is accomplished by a canonical transformation $U = e^{iS}$ where
%\begin{equation}
$S = {1 \over 2}({\bm p} \cdot \delta{\bm R} + \delta{\bm R} \cdot {\bm p}).$
%\end{equation}
The transformed Hamiltonian is
\begin{equation}
H_0 + H_1 = UHU^{-1} + i{\partial U\over \partial t} U^{-1}
\end{equation}
where $H_0$ is the same as Eq.(1) with ${\bm r}^\prime$ replaced by ${\bm r}$, and keeping only first order in $\delta{\bm R}$,
\begin{equation}
H_1 = i[S, E_0(\bm p)] + {i\over 2}\sum_\alpha
\left\{ 
\nabla_\alpha , {\partial S\over \partial t} \delta R_\alpha 
\right\}\label{H1}
\end{equation}
Next we write $\delta\bm R \sim e^{i({\bm q} \cdot {\bm r} - \omega t)}$ where $\omega = v_sq$ and $v_s$ is the sound velocity.  
We assume a slowly varying displacement, and compute the first term in Eq.(\ref{H1}) to lowest order ${\bm\nabla}\delta{\bm R}$ 
and ${\partial \delta {\bm R}\over \partial t}$.  We find
\begin{equation}
H_1 = \sum_{\alpha\beta} {\partial \delta R_\beta \over \partial r_\alpha} p_\beta
v_\alpha + i{\bm p} \cdot {\partial \delta{\bm R} \over \partial t}\label{H12}
\end{equation}
where $ v_\alpha(\bm p) = dE_0/d{\bm p}_\alpha$ is the electron velocity.  Equation~(\ref{H12}) is derived by formally expanding 
$E(\bm p)$ in power of $\bm p$.
The second term in Eq.(\ref{H12}) is of order $\omega k_F$ which is smaller than the first term by the ratio 
$\omega k_F/(q k_Fv_F) = v_s/v_F$ and can be dropped.

Now we introduce the phonons
\begin{equation}
H_{\rm ph} = \sum_{{\bm q},\lambda,\sigma} \omega_{{\bm q}\lambda} a^\dagger_{{\bm q}\lambda} a_{{\bm q}\lambda}
\end{equation}
where $\lambda$ denotes the phonon branches with polarization $\hat{\varepsilon}_{{\bm q}\lambda}$.  Expanding $\delta{\bm R}$ 
in terms of the phonon coordinate and substituting in (\ref{H12}) results in the spinon-phonon coupling term
\begin{equation}
H_{\rm s-ph} = \sum_{{\bm k},{\bm q},\lambda,\sigma}
M_{{\bm k}\lambda}({\bm q})f^\dagger_{{\bm k}+{\bm q} \sigma}
f_{{\bm k}\sigma}(a_{{\bm q}\lambda} + a^\dagger_{-{\bm q}\lambda}) ,
\end{equation}
\begin{equation}
M_{{\bm k}\lambda}(\bm q) =
({\bm k} \cdot {\hat \varepsilon}_{{\bm q}\lambda}) [{\bm q} \cdot {\bm v}({\bm k})] (2 \rho_{\rm ion}\omega_{{\bm q}\lambda})^{-1/2}\label{Mkq}
\end{equation}
where $\rho_{\rm ion}$ is the ion mass density.  The spinons are coupled to a gauge field ${\bm a}$ which 
initially has no dynamics because it was introduced to enforce the the constraint of no double occupation.  
We shall approximate $E(\bm k) = k^2/2m$ and the spinon kinetic energy is
\begin{equation}
H_{0s} = \sum_{\bm k,\sigma} {1\over 2m} ({\bm k} + {\bm a})^2 f^\dagger_{{\bm k}\sigma} f_{{\bm k}\sigma} .\label{H0s}
\end{equation}
In addition we have the impurity scattering term
$
H_{\rm imp} = \sum_{i\sigma} v_{\rm imp}({\bm r}_i) f^\dagger _\sigma({\bm r}_i)
f_\sigma ({\bm r}_i)
$
and we consider non spin flip scattering only.  We assume that impurity scattering gives an elastic scattering 
lifetime $\tau$ and mean free path $l = v_F\tau$ for the spinons.  

From this point on we can discuss the sound attenuation in spin liquids in parallel with the theory for metals 
and superconductors.  Historically the first discussion was in the hydrodynamic regime valid for $ql \ll 1$.\cite{Mason,Morse}  
The fermions are treated as a viscous medium subject to strain fields set up by the sound waves.  
This picture is clearly independent of the charge of the fermions and can directly carry over to the spinon case.  
Starting from the linearized Navier-Stokes equation, the sound wave relaxation time $\tau_s$ is
%\begin{subequations}
\begin{equation}
\tau_s^L = {1\over \rho_{\rm ion} v_s^2} \left( {4\over 3}\eta + \chi \right),  \tau_s^T = {1\over \rho_{\rm ion} v_s^2} \eta 
\end{equation}
%\end{subequations}
for the longitudinal and transverse sound, respectively, where  $\eta, \chi$ are the shear and compressional viscosities.
The sound attenuation constant $\alpha$ defined as the inverse of the phonon mean free path $l_{\rm ph}$ is 
the imaginary part of $k = (\omega/v_s)(1 + i\omega\tau_s)^{-1/2}$ and given by
%\begin{equation}
$\alpha = {\omega^2\tau_s\over 2v_s}$
%\end{equation}
in the limit $\omega\tau_s \ll 1$.  The fermion viscosity  is given by $\eta = {1\over 15} N(0) m^2 v_F^4 \tau$ 
where $N(0)$ is the density of states at the Fermi level and we can take $\chi$ to be $\ll \eta$.

The hydrodynamic theory was extended by Pippard to all values of $ql$ using a Boltzmann equation approach.\cite{Pippard}  
Pippard pointed out that when $ql \gtrsim 1$, the electrons develop local charge and current fluctuations for 
the longitudinal and transverse phonons which contribute significantly to the sound attenuation.  
Here we re-derive Pippard's results using a diagrammatic method, because it can readily be extended 
to the pairing case.  Our method is simpler than the work of Tsuneto, who combined a diagrammatic and 
Boltzmann approach. Since the diagrammatic treatment is not readily available in the literature, we provide 
the details in the supplementary material.

\begin{figure}[tbph]
\includegraphics[width=7.8cm]{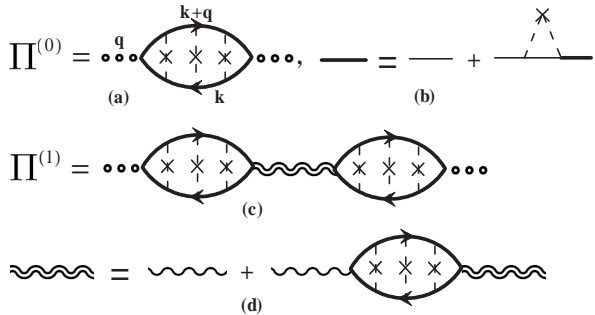}
\caption{ Feynman digrams for phonon self energy.}
\label{feyn}
\end{figure}

Let us first re-derive the results for metals.  We compute the phonon self energy $\Pi (q,\omega)$ 
due to the excitation of fermion particle-hole pairs.  It is given by the diagrams shown in Fig.1.  
The bold solid line is the fermion Green's function with self energy due to impurity scattering, 
$G_{\rm ret(adv)}(k,\omega) = (\omega-\xi_k \pm i/2\tau)^{-1}$ where $\xi_k = k^2/2m-\mu$.  The electromagnetic (EM) 
field propagation is screened by repeated bubbles representing density or current fluctuations 
for longitudinal and transverse sound respectively.  For longitudinal sound the Thomas Fermi screening 
length $k_{\rm TF}^{-1}$ is much shorter than $q^{-1}$ and it can be shown (see supp.) that the effect 
of screening is the same as calculating the unscreened bubble [Fig.1(a)] with a coupling 
matrix $\tilde{M}_{k\lambda}$ given by Eq.(\ref{Mkq}) with the trace subtracted, i.e.,
$
\tilde{M}_{{\bm k}\lambda}(\bm q) =
[
({\bm k} \cdot \hat{\varepsilon}_{{\bm q}\lambda})({\bm k} \cdot {\bm q}) - {1\over 3}k^2 ({\bm q} \cdot \hat\varepsilon_{{\bm q}\lambda})
]
 /m \sqrt{2\rho_{\rm ion}\omega_{{\bm q}\lambda}}
$.
This is the form suggested by Blount.\cite{Blount} The standard results are reproduced, and the attenuation decreases in the superconducting state 
when the quasiparticles are gapped.  In the rest of the paper we focus on the transverse sound.  
Here there is no vertex correction (impurity line across the bubbles) because the vertex given by Eq.(\ref{Mkq}) 
is odd in the $\bm k$ component along $\bm q$.  $\Pi^{(0)}$ describes the dissipation due to 
the creation of particle-hole excitation of fermions.  In $\Pi^{(1)}$ the particle-hole excitation 
creates current $\bm j$ which couples to the electromagnetic gauge field via ${\bm j}\cdot {\bm A}$.  
The gauge field acquires a self energy by coupling to current fluctuations, as shown in the third line in Fig.1.  
The resulting retarded photon propagator (double wavy line) is given by
\begin{equation}
D_{\rm ret}^{\rm EM} =
{1\over i\omega \sigma_\perp (q,\omega) + \omega^2 - c^2q^2} \label{DEM}
\end{equation}
and $c$ is the speed of light.  For ultrasound $\omega^2$ in Eq.(\ref{DEM}) can safely be ignored.  
Following Pippard's notation we write $\sigma_\perp(q,\omega) = g \sigma_0$ where $\sigma_0 = e^2n \tau/m$ 
is the DC conductivity, and
%\begin{equation}
%g = {3\over 2a}{\rm Re}
%[s_0(a) - s_2(a)]
%\end{equation}
%where
%\begin{equation}
%s_n(a) = -{i\over 2} \int^1_{-1} du {u^n\over u+i/a}
%\end{equation}
%and $a = ql/(1+i\omega\tau)$.
\begin{equation*}
g = {3\over 2a}{\rm Re}[s_2(a) - s_0(a)],\,\,
s_n(a) = {1\over 2i} \int^1_{-1} du {u^n\over u+i/a},
\end{equation*}
where $a = ql/(1+i\omega\tau)$.  
We can safely assume $\omega\tau \ll 1$ for the rest of the paper 
and set $a = ql$.  Then $g \rightarrow 1 - {2\over 15}(ql)^2$ when $ql \ll 1$ and $g \rightarrow {3\pi\over 4}(ql)^{-1}$ when $ql \gg 1$.  
We see that in the clean limit $(ql \gg 1)$ the first term in the denominator of Eq.(\ref{DEM}) is nothing 
but Landau damping $N(0)\omega / v_Fq$ while in the opposite limit $(ql \ll 1)$ it gives rise to 
the classical skin depth $k_0^{-1}$ where $k_0^2 = \omega \sigma_0 / c^2$.

The diagrams are evaluated to give (see supplementary materials for details)
\begin{equation}
{\rm Im}\Pi^{(0)}_{\rm ret} =
{\omega N(0) k_F^3 \over 4 \rho_{\rm ion} mv_s^2}
{s_1(ql) - s_3(ql) \over iql}  . \label{ImPi0}
\end{equation}
On the other hand, $\Pi_{\rm ret}^{(1)}$ is proportional to the photon propagator, and depends on 
the relative size of the inverse skin depth $k_0$, $q$ and $l^{-1}$.  Let us 
%first 
consider 
the case when $c^2q^2 \ll \omega\sigma_\perp(q,\omega)$, in which case 
$D_{\rm ret}^{\rm EM} = \left(  i\omega\sigma_\perp(q,\omega) \right)^{-1}$.  
This holds under the condition $q \ll k_0$ if $ql \ll 1$ and $q^2 \ll k_0^2 / (ql)$ if $ql \gg 1$.   
(We shall not discus the extreme clean case, $q^2 \gg k_0^2/ql$ while $q \ll k_0$, \cite{Pippard} because it is never attained for the spinon case.)
We can show that $\Pi_{\rm net}^{(1)}$ takes the remarkably simple form
\begin{equation}
\Pi^{(1)}_{\rm ret} = {1-g\over g}{\rm Im}\Pi^{(0)}_{\rm ret} + O (v_s/v_F)  .\label{Pi1}
\end{equation}
The ultrasound attenuation coefficient is given by
\begin{equation}
\alpha = -{2\over v_s} {\rm Im} (\Pi_{\rm ret}^{(0)} + \Pi_{\rm ret}^{(1)}) 
= {nm \over \rho_{\rm ion}v_s\tau} {1-g \over g} \label{alpha}
\end{equation}
where the identity $s_1 - s_3 = -{2i \over 3} (1-g) + O (v_s/v_F)$ has been used, and $n$ is the fermion density.  
Equations (\ref{alpha}) agree with Pippard's result derived using Boltzmann's equation.  Using the limit  
$g = 1 - {2\over 15}(ql)^2$ for $ql \ll 1$, 
we can verify that the hydrodynamic limit is reproduced.

Now consider the onset of superconductivity.  
%Apart from the possibility of a Hebel-Slichter type peak, 
$\Pi^{(0)}$ decreases below $T_c$ due to the opening of the energy gap (see supp.), but $\Pi^{(1)}$ is affected 
much more dramatically. Physically the onset of Meisner effect suppresses the magnetic field fluctuations 
and $\Pi^{(1)}$ drops to zero rapidly below $T_c$.  Mathematically this is because a constant term 
$e^2n_s(T)/m$ where $n_s$ is the superfluid density is added to the denominator of Eq.(\ref{DEM}) and quickly 
dominates $i\omega\sigma_\perp(q,\omega)$.  
Since $\Pi^{(1)}$ is proportional to $D^{\rm EM}_{\rm ret}$, it drops rapidly to very small value.
This is called the ``rapid fall'' and occurs over a millikelvin scale \cite{Leibowitz} in clean samples.  
We note from Eq.(\ref{Pi1}) that the fractional size of the drop 
is $(1-g)$ which is very small ($ \sim {2\over 15} (ql)^2$) for $ql \ll 1$ but almost 
unity ($1- {3\pi \over 4 ql}$) in the clean limit of $ql \gg 1$.  
%In extremely clean samples, $k^2_0 / (ql)$ can become larger than $q^2$ even if $q \ll k_0$ and 
%the $c^2q^2$ term in Eq.(\ref{DEM}) needs to be kept, leading to more complicated formulas.\cite{Pippard}  
%We shall see that this limit is never attained for the spinon case and we shall not discuss this further.

Next we turn our attention to the attenuation of transverse sound by spinons.  
The main difference is that the spinons and gauge fields are treated in 2D.  
Furthermore, the Maxwell term $\omega^2 - c^2q^2$ is missing in the photon propagator.  
The dynamics of the gauge field is generated by spinon current fluctuation and instead of Eq.(\ref{DEM}) we have
\begin{equation}
D^T_{\rm ret} = {1\over i\omega \tilde{\sigma}_\perp (q,\omega) - \chi q^2} \label{DT}
\end{equation}
where $\tilde{\sigma}_\perp = \tilde{g} \tilde{\sigma}_0$, $\tilde{\sigma}_0 = n \tau/m$, 
$n$ is the spinon density and $\chi = 1/(24 \pi m)$ is the Landau diamagnetism.\cite{Lee/Nagaosa}  
Note that according to Eq.(\ref{H0s}) the coupling constant to the gauge field has been set to unity instead of $e$.  
The factor $\tilde{g}$ is calculated in 2D and is given by
%\begin{eqnarray}
%\tilde{g} &=& -{2 \over a} [t_0(a) - t_2(a)] \nonumber , \,\,\, {\rm where} \\
%t_n(a) &=&  -{i\over 2} \int^{2\pi}_0 d\theta  {\cos^n\theta \over \cos\theta + i/a}  .
%\end{eqnarray}
\begin{equation*}
\tilde{g} = {2 \over a} [t_2(a) - t_0(a)], \,\,
t_n(a) =  {1\over 2i} \int^{2\pi}_0  d\theta {\cos^n\theta \over \cos\theta + i/a}  .
\end{equation*}
Once again, $\tilde{g}$ can be considered a function of $ql$, $\tilde{g}=1-{(ql)^2\over 4}$ for $ql\ll 1$
and $\tilde{g}={2\over ql}$ for $ql\gg 1$.
%and approaches 1 and $(ql)^{-1}$ 
%for $ql$ less than or greater than unity, respectively.

\begin{figure}[tbph]
\includegraphics[width=6.0cm]{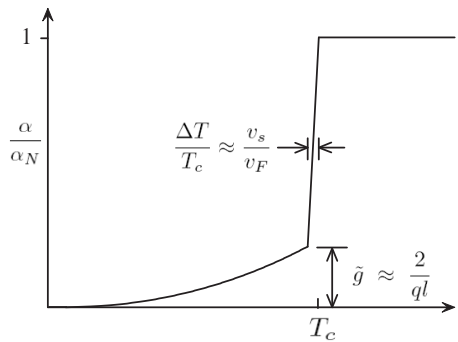}
\caption{Schematic plot of the attenuation of transverse sound (for $ql \gtrsim 1)$ in a spin liquid normalized to $\alpha_N$, the value in the Fermi liquid state.  The ``rapid fall'' over a narrow temperature range $\Delta T$ at the transition to the spinon paired state is due to the Meissner effect of the gauge field.}
\label{fall}
\end{figure}

Just as in the EM case, we consider the case when $\chi q^2 \ll \omega\tilde{\sigma}_\perp$, 
i.e., $D^T_{\rm ret} = (i\omega\tilde{\sigma}_\perp)^{-1}$ and we conclude that $\alpha$ 
is given by Eq.(\ref{alpha}) with g replaced by $\tilde{g}$. In the Fermi liquid state, we can use this formula to get information on the spinon mass and lifetime $\tau$ by studying the $q$ (i.e., $\omega$) dependence.  At low temperature $\tau$ is a constant dominated by disorder \cite{MYamashita10} while at finite temperature $\tau$ may have interesting $T$ dependence due to scattering by other spinons or phonons. If the spinons are paired, the rapid fall 
also occurs just below $T_c$.  Next we show that 
%this
the condition $\chi q^2 \ll \omega\tilde{\sigma}_\perp$ mentioned above  
is the only relevant limit.  For $ql \gtrsim 1$, $\omega\tilde{\sigma}_\perp$ 
is estimated to be $(v_s/v_F) k_F^2/m$ and will dominate $\chi q^2$ as long as $v_s/v_F \gg (q/k_F)^2$.  
Since $v_s/v_F \approx 10^{-3}$, this condition is satisfied for most accessible ultrasound frequencies.  
In the opposite case $(ql \ll 1)$ the condition is $(v_s/v_F)ql >> (q/k_F)^2$ and is easier to violate.  
However, the rapid fall is very small in this case and difficult to detect and of little interest to us.

Finally we can estimate the temperature range $\Delta T = T_c-T$ of the rapid fall as sketched in Fig.2.  
Since $i\omega\tilde{\sigma}_\perp$ is replaced by 
$i\omega\tilde{\sigma} - n_s(T)/m$ in Eq.(\ref{DT}) below $T_c$ we find that 
%$\Pi^{(1)}_s(T) = [\omega\tilde{\sigma}_0 \tilde{g} m / n_s(T)]\Pi^{(1)}_N$ where $\Pi^{(1)}_N$ and $\Pi^{(0)}_N$ 
%are the values in the normal state.  
$\rm{Im} \Pi_s^{(1)} =\rm{Im} \Pi^{(1)}_N/\left(1 + (n_s(T)/m\omega\tilde{\sigma_0}\tilde{g})^2 \right)$ 
where $\Pi^{(1)}_{s,N}$ are the values in the superconducting and normal states.  For $ql \gg 1$, 
attenuation is dominated by $\Pi^{(1)}$ and we estimate $\Delta T$ as the temperature when $\rm{Im} \Pi_s^{(1)}$ 
has dropped half the normal state value. 
We assume the mean field (BCS) behavior $n_s(T) = 2n (\Delta T/T_c)$ and  
%For $ql \gg 1$, $\Pi_N^{(0)}\approx 2\Pi^{(1)}_N/ql$ and we estimate $\Delta T$ by setting $\Pi_s^{(1)}(T) = \Pi^{(0)}_N$.  
%Using Eq.(\ref{Pi1}) 
we find
\begin{equation}
{\Delta T\over T_c} \approx {v_s\over v_F}  .\label{deltaT}
\end{equation}

We conclude that fermionic spinons 
in a fermion 
couple to phonons in a way which is identical to electron phonon coupling 
in the long wavelength limit.  For $ql \gg 1$ the attenuation of transverse ultrasound is dominated by 
a component which is due to the fluctuations of transverse gauge fields.  At the pairing transition of the spinon, 
this component is suppressed by the onset of the Meissner effect for the gauge field, leading to a rapid drop in 
attenuation in a very narrow temperature range below $T_c$ given by Eq.(\ref{deltaT}).  After the rapid fall, 
the attenuation is reduced in the usual way by the gapping of the quasiparticles.  We believe the phenomenon of 
the rapid fall in the attenuation of transverse ultrasound gives a clear signature of spinon pairing and 
the existence of $U(1)$ gauge fields.

We thank T. Senthil for helpful comments.  PAL acknowledges support by NSF under DMR--0804040. YZ is supported by NSFC (No.11074218),
973 Program (No.2011CBA00103), and the Fundamental Research Funds for the Central Universities in China.

\clearpage

\begin{widetext}

\section{Supplementary Material}

\subsection{Sound absorption and attenuation in a liquid}

Fluid viscosity will cause sound absorption and attenuation in a liquid. To
discuss this effect, we begin with linearlized Navier-Stokes equation, 
\begin{equation}
\rho \frac{\partial \mathbf{u}}{\partial t}=-\nabla p+\left( \frac{4}{3}\eta
+\chi \right) \nabla \left( \nabla \cdot \mathbf{u}\right) -\eta \nabla
\times \nabla \times \mathbf{u,}  \label{LNSE}
\end{equation}%
where $\rho $ is the fluid density, $\mathbf{u}$ is the flow velocity, $p$
is the pressure, $\eta $ and $\chi $ are the shear and compressional
viscosities respectivley. The instantaneous density $\rho $ may be further
written as $\rho =\rho _{0}\left( 1+s\right) $, where $s$ is a small
fraction and $\rho _{0}$ is a constant. Then equation of coninuity has the
following form 
\begin{equation}
\nabla \cdot \mathbf{u}=-\frac{\partial s}{\partial t}.
\end{equation}%
The acoustic pressure $p$ is found to be of the form%
\begin{equation}
p=\left( \frac{\partial p}{\partial \rho }\right) _{\rho _{0}}\rho
_{0}s=\rho _{0}v_{s}^{2}s
\end{equation}%
in terms of $s$ and the sound velocity $v_{s}$. So that we have a lossy wave
equation 
\begin{equation}
\left( 1+\tau _{s}\frac{\partial }{\partial t}\right) \nabla ^{2}p=\frac{1}{%
v_{s}^{2}}\frac{\partial ^{2}p}{\partial t^{2}},
\end{equation}%
with a relaxation time given by Eqs.(10). If we assume monofrequency motion,
the above wave quations is reduced to a lossy Helmholtz equation%
\begin{equation}
\nabla ^{2}p+k^{2}p=0,
\end{equation}%
where $k=\frac{\omega }{v_{s}}\frac{1}{\left( 1+i\omega \tau _{s}\right)
^{1/2}}$. The solution for $\alpha =-$Im$k$ reads 
\begin{equation}
\alpha =\frac{1}{\sqrt{2}}\frac{\omega }{v_{s}}\left[ \frac{\sqrt{1+\left(
\omega \tau _{s}\right) ^{2}}-1}{1+\left( \omega \tau _{s}\right) ^{2}}%
\right] ^{1/2}.
\end{equation}%
In the limit $\omega \tau _{s}\ll 1$, we have the sound attenuation
coefficient in Eq.(11).

\subsection{Screened spinon phonon coupling for longitudinal mode}

Without the loss of generality, the spinon phonon coupling $M_{\mathbf{k}%
\lambda }\left( \mathbf{q}\right) $ for longitudinal sound can be written in
terms of spherical harmonic functions,%
\begin{equation}
M_{\mathbf{k}\lambda }\left( \mathbf{q}\right) =\frac{1}{3}%
f(k,q)+\sum\limits_{l\geq 1,m}a_{lm}Y_{lm}\left( \theta ,\phi \right) ,
\label{MkE}
\end{equation}%
where $f(k,q)=\frac{k^{2}q}{m\sqrt{2\rho _{ion}\omega _{\mathbf{q}\lambda }}}
$, $\theta $ is the angle between $\mathbf{k}$ and $\mathbf{q}$, $\phi $ is
the azimuthal angle of $\mathbf{k}$. For longitudinal mode, the Thomas Fermi
screening length $k_{\text{TF}}^{-1}$ is much shorter than $q^{-1}$. We
shall show that the monople part $\frac{1}{3}f(k,q)$ will be screened by
charge fluctuation, resulting in an effective coupling matrix $\tilde{M}_{%
\mathbf{k}\lambda }\left( \mathbf{q},\omega \right) $ containing only higher
order terms in multipole expansion (\ref{MkE}) and a term of order of $%
v_{s}/v_{F}$.

\begin{figure}[hpbt]
\includegraphics[width=7.6cm]{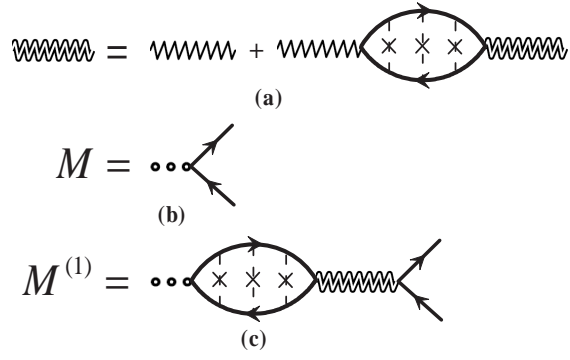}
\caption{(a) Screening effect of spinon density-density interaction $U$,
zigzag lines denote unscreened interaction $U^{(0)}$, double zigzag lines
denote screened interaction $U$, the bubble is for the polarization function 
$P$. (b) The bare spinon phonon coupling. (c) Screened part of spinon phonon
coupling. }
\label{fig3}
\end{figure}

We firstly consider screening effect of the bare spinon density-density
interaction $U^{(0)}\left( \mathbf{q},\omega \right) $, which gives rise to
the screened interaction%
\begin{equation}
U\left( \mathbf{q},\omega \right) =\frac{U^{\left( 0\right) }\left( \mathbf{q%
},\omega \right) }{1-P\left( \mathbf{q},\omega \right) U^{\left( 0\right)
}\left( \mathbf{q},\omega \right) },
\end{equation}%
where $P$ is the polarization function in an impure fermi system which is
given by the bubble diagram in Fig.3(a). The screening of spinon phonon
coupling is very similar to that of $U$. These two can be evaluated through
the same vertex function. We divide the screened spinon phonon coupling $%
\tilde{M}_{\mathbf{k}\lambda }\left( \mathbf{q},\omega \right) $ into two
parts,%
\begin{equation}
\tilde{M}_{\mathbf{k}\lambda }\left( \mathbf{q},\omega \right) =M_{\mathbf{k}%
\lambda }\left( \mathbf{q}\right) +M_{\mathbf{k}\lambda }^{(1)}\left( 
\mathbf{q},\omega \right) ,
\end{equation}%
where $M_{\mathbf{k}\lambda }^{(1)}$ in Matsubara frequency is given by 
\begin{equation}
M_{\mathbf{k}\lambda }^{(1)}\left( \mathbf{q},i\omega _{n}\right) =\frac{2}{%
\beta }\sum_{ik_{m},\mathbf{k}}M_{\mathbf{k}\lambda }\left( \mathbf{q}%
\right) \mathcal{G}\left( \mathbf{k+q},ik_{m}\right) \mathcal{G}\left( 
\mathbf{k},i\omega _{n}+ik_{m}\right) U\left( \mathbf{q},i\omega _{n}\right)
\gamma \left( \mathbf{q},\mathbf{k},ik_{m},i\omega _{n}+ik_{m}\right) ,
\end{equation}%
as shown in Fig.3(c). The vertex function $\gamma $ is the solution of the
following equation,%
\begin{eqnarray}
\gamma \left( \mathbf{q,k},ik_{m},i\omega _{n}+ik_{m}\right)
&=&1+n_{imp}\sum_{\mathbf{k}^{\prime }}\mathcal{G}\left( \mathbf{k}^{\prime }%
\mathbf{+q},ik_{m}\right) \mathcal{G}\left( \mathbf{k}^{\prime },i\omega
_{n}+ik_{m}\right)  \notag \\
&&\times T_{\mathbf{k+q,k}^{\prime }\mathbf{+q}}\left( ik_{m}\right) T_{%
\mathbf{k}^{\prime }\mathbf{k}}\left( i\omega _{n}+ik_{m}\right) \gamma
\left( \mathbf{q,k}^{\prime },ik_{m},i\omega _{n}+ik_{m}\right) ,
\end{eqnarray}%
where $T_{\mathbf{k}^{\prime }\mathbf{k}}$ is the transition matrix due to
multiple impurity scattering and $n_{imp}$ is the impurity density.

To proceed we shall make use of the following approximations,%
\begin{eqnarray}
\mathcal{G}_{\text{adv}}\left( \mathbf{k+q},\epsilon \right) \mathcal{G}_{%
\text{ret}}\left( \mathbf{k},\epsilon +\omega \right) &\simeq &\frac{2i\pi
\delta \left( \epsilon -\xi _{\mathbf{k}}\right) }{qv_{F}\cos \theta +\omega
+i/\tau }, \\
\int \frac{d\epsilon }{2\pi i}n_{F}\left( \epsilon \right) \mathcal{G}_{%
\text{ret}}\left( \mathbf{k+q},\epsilon \right) \mathcal{G}_{\text{ret}%
}\left( \mathbf{k},\epsilon +\omega \right) &\simeq &-\frac{\partial
n_{F}\left( \xi _{\mathbf{k}}\right) }{\partial \xi _{\mathbf{k}}}, \\
\int \frac{d\epsilon }{2\pi i}n_{F}\left( \epsilon \right) \mathcal{G}_{%
\text{adv}}\left( \mathbf{k+q},\epsilon \right) \mathcal{G}_{\text{adv}%
}\left( \mathbf{k},\epsilon -\omega \right) &\simeq &\frac{\partial
n_{F}\left( \xi _{\mathbf{k}}\right) }{\partial \xi _{\mathbf{k}}}.
\end{eqnarray}%
For $l\geq 1$, one sees that%
\begin{equation}
\int d\Omega Y_{lm}\left( \theta ,\phi \right) \int \frac{d\epsilon }{2\pi i}%
n_{F}\left( \epsilon \right) \mathcal{G}_{\text{ret(adv)}}\left( \mathbf{k+q}%
,\epsilon \right) \mathcal{G}_{\text{ret(adv)}}\left( \mathbf{k},\epsilon
+\omega \right) =0.
\end{equation}%
With the help of optical sum rule and using the relation 
\begin{equation*}
\frac{1}{2\tau }=-n_{imp}\text{Im}T_{\mathbf{kk}},
\end{equation*}%
we find that%
\begin{eqnarray}
\gamma _{1}\left( \mathbf{q,k},\epsilon ,\epsilon +\omega \right) &\equiv
&\gamma \left( \mathbf{q,k},\epsilon -i\eta ,\epsilon +\omega +i\eta \right)
=\frac{1}{1+s_{0}\left( a\right) /a}, \\
\gamma _{2}\left( \mathbf{q,k},\epsilon ,\epsilon +\omega \right) &\equiv
&\gamma \left( \mathbf{q,k},\epsilon +i\eta ,\epsilon +\omega +i\eta \right)
=1,
\end{eqnarray}%
where $a=ql/(1+i\omega \tau )$. Hereafter we shall assume $\omega \tau =%
\frac{v_{s}}{v_{F}}ql\ll 1$ and set $a=ql$. So that the polarization
function $P$ can be evaluated as follows,%
\begin{equation}
P\left( \mathbf{q},i\omega _{n}\right) =\sum_{\mathbf{k}}\sum_{ik_{m}}%
\mathcal{G}\left( \mathbf{k+q},ik_{m}\right) \mathcal{G}\left( \mathbf{k}%
,i\omega _{n}+ik_{m}\right) \gamma \left( \mathbf{q,k},ik_{m},i\omega
_{n}+ik_{m}\right) ,
\end{equation}%
Taking analytical continuation, we have%
\begin{eqnarray}
P\left( \mathbf{q},\omega \right) &=&2i\sum_{\mathbf{k}}\int_{-\infty
}^{\infty }\frac{d\epsilon }{2\pi }n_{F}\left( \epsilon \right) \gamma
_{2}\left( \mathbf{q,k},\epsilon ,\epsilon +\omega \right)  \notag \\
&&\times \left[ \mathcal{G}_{\text{ret}}\left( \mathbf{k},\epsilon +\omega
\right) \mathcal{G}_{\text{ret}}\left( \mathbf{k+q},\epsilon \right) -%
\mathcal{G}_{\text{adv}}\left( \mathbf{k+q},\epsilon -\omega \right) 
\mathcal{G}_{\text{adv}}\left( \mathbf{k},\epsilon \right) \right]  \notag \\
&&+2i\sum_{\mathbf{k}}\int_{-\infty }^{\infty }\frac{d\epsilon }{2\pi }%
n_{F}\left( \epsilon \right) \gamma _{1}\left( \mathbf{q,k},\epsilon
,\epsilon +\omega \right)  \notag \\
&&\times \left[ -\mathcal{G}_{\text{ret}}\left( \mathbf{k},\epsilon +\omega
\right) \mathcal{G}_{\text{adv}}\left( \mathbf{k+q},\epsilon \right) +%
\mathcal{G}_{\text{adv}}\left( \mathbf{k+q},\epsilon -\omega \right) 
\mathcal{G}_{\text{ret}}\left( \mathbf{k},\epsilon \right) \right]  \notag \\
&=&-N\left( 0\right) \left[ 1-i\omega \tau \frac{s_{0}\left( a\right) /a}{%
1+s_{0}\left( a\right) /a}\right] ,
\end{eqnarray}%
When the Thomas Fermi screening length $k_{\text{TF}}^{-1}\ll q^{-1}$, $%
N\left( 0\right) U^{\left( 0\right) }\left( \mathbf{q},\omega \right) \gg 1$%
, $U\left( \mathbf{q},\omega \right) $ can be reduced to%
\begin{equation}
U\left( \mathbf{q},\omega \right) =\frac{1}{N\left( 0\right) \left[
1-i\omega \tau \frac{s_{0}\left( a\right) /a}{1+s_{0}\left( a\right) /a}%
\right] }.
\end{equation}%
Then we are ready to calculate $M_{\mathbf{k}\lambda }^{(1)}\left( \mathbf{q}%
\right) $,%
\begin{eqnarray}
M_{\mathbf{k}\lambda }^{(1)}\left( \mathbf{q},\omega \right) &=&\frac{2}{%
\beta }\sum_{\mathbf{k}}M_{\mathbf{k}\lambda }\left( \mathbf{q}\right)
\sum_{ik_{m}}\mathcal{G}\left( \mathbf{k+q},ik_{m}\right) \mathcal{G}\left( 
\mathbf{k},i\omega _{n}+ik_{m}\right) \gamma U\left( \mathbf{q},i\omega
_{n}\right)  \notag \\
&=&\frac{1}{3}f\left( k,q\right) P\left( \mathbf{q},\omega \right) U\left( 
\mathbf{q},\omega \right) -2i\omega f\left( k,q\right) \sum_{\mathbf{k}%
}\left( \cos ^{2}\theta -\frac{1}{3}\right)  \notag \\
&&\times \int_{-\infty }^{\infty }\frac{d\epsilon }{2\pi }\left[ -\frac{%
\partial n_{F}\left( \epsilon \right) }{\partial \epsilon }\right] \mathcal{G%
}_{\text{ret}}\left( \mathbf{k},\epsilon +\omega \right) \mathcal{G}_{\text{%
adv}}\left( \mathbf{k+q},\epsilon \right) \gamma _{1}U\left( \mathbf{q}%
,\omega \right)  \notag \\
&=&f\left( k,q\right) \left[ -\frac{1}{3}+i\frac{\omega \tau }{a}\frac{%
s_{2}\left( a\right) -s_{0}\left( a\right) /3}{1+\left( 1-i\omega \tau
\right) s_{0}\left( a\right) /a}\right]
\end{eqnarray}%
Therefore%
\begin{equation*}
\tilde{M}_{\mathbf{k}\lambda }\left( \mathbf{q}\right) =\frac{k^{2}q}{m\sqrt{%
2\rho _{ion}\omega _{\mathbf{q}\lambda }}}\left[ \cos ^{2}\theta -\frac{1}{3}%
+i\frac{v_{s}}{v_{F}}\frac{s_{2}\left( a\right) -s_{0}\left( a\right) /3}{%
1+\left( 1-i\omega \tau \right) s_{0}\left( a\right) /a}\right] .
\end{equation*}%
Thus we can calculate the sound attenuation constant through the unscreened
bubble [Fig1.(a)] with the traceless coupling matrix,%
\begin{equation}
\tilde{M}_{\mathbf{k}\lambda }\left( \mathbf{q}\right) =\frac{\left( \mathbf{%
k}\cdot \mathbf{q}\right) \left( \mathbf{k}\cdot \hat{\varepsilon}_{\mathbf{q%
}\lambda }\right) -\frac{1}{3}k^{2}\left( \mathbf{q}\cdot \hat{\varepsilon}_{%
\mathbf{q}\lambda }\right) }{m\sqrt{2\rho _{ion}\omega _{\mathbf{q}\lambda }}%
}+O\left( \frac{v_{s}}{v_{F}}\right) ,
\end{equation}%
for longitudinal mode.

\subsection{Longitudinal sound attenuation}

Then we calculate longitudinal sound attenuation constant using the above
spinon phonon coupling. In this case, we can express $\Pi \left( \mathbf{q}%
,i\omega _{n}\right) $ in terms of a vertex function $\Gamma \left( \mathbf{%
q,k},ik_{m},i\omega _{n}+ik_{m}\right) $, 
\begin{equation}
\Pi \left( \mathbf{q},i\omega _{n}\right) =\frac{2}{\beta }\sum_{\mathbf{k}}%
\tilde{M}_{\mathbf{k}\lambda }\left( \mathbf{q}\right) \sum_{ik_{m}}\mathcal{%
G}\left( \mathbf{k+q},ik_{m}\right) \mathcal{G}\left( \mathbf{k},i\omega
_{n}+ik_{m}\right) \Gamma \left( \mathbf{q,k},ik_{m},i\omega
_{n}+ik_{m}\right) ,  \label{PiL1}
\end{equation}%
where the vertex function $\Gamma $ obeys%
\begin{eqnarray}
\Gamma \left( \mathbf{q,k},ik_{m},i\omega _{n}+ik_{m}\right) &=&\tilde{M}_{%
\mathbf{k}\lambda }\left( \mathbf{q}\right) +n_{imp}\sum_{\mathbf{k}^{\prime
}}\mathcal{G}\left( \mathbf{k}^{\prime }\mathbf{+q},ik_{m}\right) \mathcal{G}%
\left( \mathbf{k}^{\prime },i\omega _{n}+ik_{m}\right)  \notag \\
&&\times T_{\mathbf{k+q,k}^{\prime }\mathbf{+q}}\left( ik_{m}\right) T_{%
\mathbf{k}^{\prime }\mathbf{k}}\left( i\omega _{n}+ik_{m}\right) \Gamma
\left( \mathbf{q,k}^{\prime },ik_{m},i\omega _{n}+ik_{m}\right) ,
\end{eqnarray}%
and is shown in Fig.4 diagramatically.

\begin{figure}[tbph]
\includegraphics[width=7.5cm]{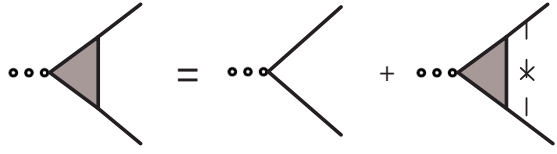}
\caption{Vertex correction in longitudinal sound, the dash line across the
vertex denotes impurity scattering.}
\label{fig4}
\end{figure}

It is more convenient to write the analytical continuations of $\Gamma $ as
follows, 
\begin{eqnarray}
\Gamma _{1}\left( \mathbf{q,k},\epsilon ,\epsilon +\omega \right) &\equiv
&\Gamma \left( \mathbf{q,k},\epsilon -i\eta ,\epsilon +\omega +i\eta \right)
, \\
\Gamma _{2}\left( \mathbf{q,k},\epsilon ,\epsilon +\omega \right) &\equiv
&\Gamma \left( \mathbf{q,k},\epsilon +i\eta ,\epsilon +\omega +i\eta \right)
.
\end{eqnarray}%
We find out the following two solutions,%
\begin{eqnarray}
\Gamma _{1}\left( \mathbf{q,k},\epsilon ,\epsilon +\omega \right) &=&\tilde{M%
}_{\mathbf{k}\lambda }\left( \mathbf{q}\right) -\frac{qk_{F}^{2}}{m\sqrt{%
2\rho _{ion}\omega _{\mathbf{q}\lambda }}}\frac{\left[ s_{2}\left( a\right) -%
\frac{1}{3}s_{0}\left( a\right) \right] /a}{1+s_{0}\left( a\right) /a}, \\
\Gamma _{2}\left( \mathbf{q,k},\epsilon ,\epsilon +\omega \right) &=&\tilde{M%
}_{\mathbf{k}\lambda }\left( \mathbf{q}\right) .
\end{eqnarray}%
Thus the retarded phonon polarization function can be calculated as%
\begin{eqnarray*}
\Pi _{\text{ret}}\left( \mathbf{q},\omega \right) &=&2i\sum_{\mathbf{k}}%
\tilde{M}_{\mathbf{k}\lambda }\left( \mathbf{q}\right) \int_{-\infty
}^{\infty }\frac{d\epsilon }{2\pi }n_{F}\left( \epsilon \right) \left[ 
\mathcal{G}_{\text{ret}}\left( \mathbf{k},\epsilon +\omega \right) \mathcal{G%
}_{\text{ret}}\left( \mathbf{k+q},\epsilon \right) \Gamma _{2}\left( \mathbf{%
q,k},\epsilon ,\epsilon +\omega \right) \right. \\
&&\left. -\mathcal{G}_{\text{adv}}\left( \mathbf{k+q},\epsilon -\omega
\right) \mathcal{G}_{\text{adv}}\left( \mathbf{k},\epsilon \right) \Gamma
_{2}^{\ast }\left( \mathbf{q,k},\epsilon ,\epsilon +\omega \right) \right] \\
&&+2i\sum_{\mathbf{k}}\tilde{M}_{\mathbf{k}\lambda }\left( \mathbf{q}\right)
\int_{-\infty }^{\infty }\frac{d\epsilon }{2\pi }n_{F}\left( \epsilon
\right) \Gamma _{1}\left( \mathbf{q,k},\epsilon ,\epsilon +\omega \right) \\
&&\times \left[ -\mathcal{G}_{\text{ret}}\left( \mathbf{k},\epsilon +\omega
\right) \mathcal{G}_{\text{adv}}\left( \mathbf{k+q},\epsilon \right) +%
\mathcal{G}_{\text{adv}}\left( \mathbf{k+q},\epsilon -\omega \right) 
\mathcal{G}_{\text{ret}}\left( \mathbf{k},\epsilon \right) \right] .
\end{eqnarray*}%
The imaginary part of the terms with $\Gamma _{2}$ and $\Gamma _{2}^{\ast }$
will vanish, resulting in%
\begin{eqnarray}
\text{Im}\Pi _{\text{ret}}\left( \mathbf{q},\omega \right) &=&-2\omega \text{%
Re}\sum_{\mathbf{k}}\tilde{M}_{\mathbf{k}\lambda }\left( \mathbf{q}\right)
\int_{-\infty }^{\infty }\frac{d\epsilon }{2\pi }\left[ -\frac{\partial
n_{F}\left( \epsilon \right) }{\partial \epsilon }\right] \mathcal{G}_{\text{%
ret}}\left( \mathbf{k},\epsilon +\omega \right) \mathcal{G}_{\text{adv}%
}\left( \mathbf{k+q},\epsilon \right) \Gamma _{1}\left( \mathbf{q,k}%
,\epsilon ,\epsilon +\omega \right)  \notag \\
&=&\frac{\omega N\left( 0\right) k_{F}^{3}}{2m\rho _{ion}v_{s}}\text{Re}%
\left\{ s_{4}\left( a\right) -\frac{2}{3}s_{2}\left( a\right) +\frac{1}{9}%
s_{0}\left( a\right) -\frac{\left[ s_{2}\left( a\right) -\frac{1}{3}%
s_{0}\left( a\right) \right] ^{2}}{a+s_{0}\left( a\right) }\right\} ,
\end{eqnarray}
Using the expressions of $s_{n}\left( a\right) $,%
\begin{eqnarray*}
s_{0}\left( a\right) &=&-\tan ^{-1}a, \\
s_{2}\left( a\right) &=&-\frac{a-\tan ^{-1}a}{a^{2}}, \\
s_{4}\left( a\right) &=&-\frac{-3a+a^{3}+3\tan ^{-1}a}{3a^{4}},
\end{eqnarray*}%
we obtain the longitudinal sound attenuation constant,%
\begin{eqnarray}
\alpha &=&\frac{\omega N\left( 0\right) k_{F}^{3}}{m\rho _{ion}v_{s}^{2}}%
\frac{1}{3ql}\left[ \frac{q^{2}l^{2}\tan ^{-1}\left( ql\right) }{ql-\tan
^{-1}\left( ql\right) }-\frac{1}{3}\right]  \notag \\
&=&\frac{nm}{\rho _{ion}v_{s}\tau }\left[ \frac{q^{2}l^{2}\tan ^{-1}\left(
ql\right) }{ql-\tan ^{-1}\left( ql\right) }-\frac{1}{3}\right] ,  \label{aL}
\end{eqnarray}%
where the density of spinons $n=\frac{N\left( 0\right) k_{F}^{2}}{3m}$ is
used. Eq.(\ref{aL}) coincides with Pippard's longitudinal sound attenuation
constant.

Similarly, we obtain the longitudinal sound attenuation constant in 2D by
replacing $s_{n}\left( a\right) $ by $t_{n}\left( a\right) $ so on and so
forth,%
\begin{eqnarray}
\alpha &=&-\frac{\omega N\left( 0\right) k_{F}^{3}}{m\rho _{ion}v_{s}^{2}}%
\text{Re}\left\{ t_{4}\left( a\right) -t_{2}\left( a\right) +\frac{1}{4}%
t_{0}\left( a\right) -\frac{\left[ t_{2}\left( a\right) -\frac{1}{2}%
t_{0}\left( a\right) \right] ^{2}}{a+t_{0}\left( a\right) }\right\}  \notag
\\
&=&\frac{nm}{\rho _{ion}v_{s}\tau }\left[ \frac{q^{2}l^{2}}{1-\sqrt{%
1+q^{2}l^{2}}}-\frac{1}{2}\right] .  \label{aL2D}
\end{eqnarray}

\subsection{Transverse sound attenuation in an electron gas}

We proceed with transverse sound, where there is no vertex correction
(impurity line across the bubbles) because the vertex is odd in the $\mathbf{%
k}$ component along $\mathbf{q}$. The single bubble diagram $\Pi ^{(0)}$ can
be evaluated as follows,%
\begin{eqnarray}
\Pi _{\text{ret}}^{\left( 0\right) }\left( \mathbf{q},\omega \right)
&=&-2i\sum_{\mathbf{k}}[M_{\mathbf{k}\lambda }\left( \mathbf{q}\right)
]^{2}\int_{-\infty }^{\infty }\frac{d\epsilon }{2\pi }\left\{ \left[
n_{F}\left( \epsilon \right) -n_{F}\left( \epsilon +\omega \right) \right] 
\mathcal{G}_{\text{ret}}\left( \mathbf{k},\epsilon +\omega \right) \mathcal{G%
}_{\text{adv}}\left( \mathbf{k+q},\epsilon \right) \right.  \notag \\
&&\left. -n_{F}\left( \epsilon \right) [\mathcal{G}_{\text{ret}}\left( 
\mathbf{k},\epsilon +\omega \right) \mathcal{G}_{\text{ret}}\left( \mathbf{%
k+q},\epsilon \right) -\mathcal{G}_{\text{adv}}\left( \mathbf{k+q},\epsilon
-\omega \right) \mathcal{G}_{\text{adv}}\left( \mathbf{k},\epsilon \right)
]\right\} .  \label{PiT01}
\end{eqnarray}%
Similarly to longitudinal mode, the imaginary part of $\mathcal{G}_{\text{ret%
}}\mathcal{G}_{\text{ret}}$ and $\mathcal{G}_{\text{adv}}\mathcal{G}_{\text{%
adv}}$ terms vanishes. So that%
\begin{eqnarray}
\text{Im}\Pi _{\text{ret}}^{\left( 0\right) }\left( \mathbf{q},\omega
\right) &=&-2\omega \text{Re}\sum_{\mathbf{k}}[M_{\mathbf{k}\lambda }\left( 
\mathbf{q}\right) ]^{2}\int_{-\infty }^{\infty }\frac{d\epsilon }{2\pi }%
\left[ -\frac{\partial n_{F}\left( \epsilon \right) }{\partial \epsilon }%
\right] \mathcal{G}_{\text{ret}}\left( \mathbf{k},\epsilon +\omega \right) 
\mathcal{G}_{\text{adv}}\left( \mathbf{k+q},\epsilon \right)  \notag \\
&=&\frac{\omega N\left( 0\right) k_{F}^{3}}{4\rho _{ion}mv_{s}}\text{Re}%
\left[ s_{2}\left( a\right) -s_{4}\left( a\right) \right]  \notag \\
&=&\frac{\omega N\left( 0\right) k_{F}^{3}}{4\rho _{ion}mv_{s}}\frac{%
s_{1}\left( ql\right) -s_{3}\left( ql\right) }{iql},  \label{PiT02}
\end{eqnarray}%
where the relation%
\begin{equation}
s_{n}\left( a\right) =-\frac{i\left[ 1-\left( -\right) ^{n}\right] }{2n}-%
\frac{i}{a}s_{n-1}\left( a\right)  \label{itsn}
\end{equation}%
is used to derive the last line in Eq.(\ref{PiT02}) and we keep all the
terms up to the leading order of $v_{s}/v_{F}$.

Now we go on calculating $\Pi ^{(1)}$. Since the gauge field propagator is
of the form%
\begin{equation}
D_{\alpha \beta }\left( \mathbf{q},i\omega _{n}\right) =\left( \delta
_{\alpha \beta }-q_{\alpha }q_{\beta }/q^{2}\right) D^{T}\left( \mathbf{q}%
,i\omega _{n}\right) ,
\end{equation}%
$\Pi ^{\left( 1\right) }\left( \mathbf{q},i\omega _{n}\right) $ can be
written as, 
\begin{eqnarray}
\Pi ^{\left( 1\right) }\left( \mathbf{q},i\omega _{n}\right) &=&\frac{e^{2}}{%
\beta ^{2}m^{2}}\sum_{\alpha \beta }\sum_{\mathbf{k}}M_{\mathbf{k}\lambda
}\left( \mathbf{q}\right) \sum_{ik_{m}}\mathcal{G}\left( \mathbf{k+q}%
,ik_{m}\right) \mathcal{G}\left( \mathbf{k},i\omega _{n}+ik_{m}\right) 
\notag \\
&&\times \sum_{\mathbf{k}^{\prime }}M_{\mathbf{k}^{\prime }\lambda }\left( 
\mathbf{q}\right) \left( 2k_{\alpha }+q_{\alpha }\right) \left( 2k_{\beta
}^{\prime }+q_{\beta }\right) \left( \delta _{\alpha \beta }-q_{\alpha
}q_{\beta }/q^{2}\right) D^{T}\left( \mathbf{q},i\omega _{n}\right)  \notag
\\
&&\times \sum_{ik_{m}^{\prime }}\mathcal{G}\left( \mathbf{k}^{\prime }%
\mathbf{+q},ik_{m}^{\prime }\right) \mathcal{G}\left( \mathbf{k}^{\prime
},i\omega _{n}+ik_{m}^{\prime }\right)  \notag \\
&=&\frac{4e^{2}}{\beta ^{2}m^{2}}D^{T}\left( \mathbf{q},i\omega _{n}\right)
\sum_{\mathbf{kk}^{\prime }}M_{\mathbf{k}\lambda }\left( \mathbf{q}\right)
M_{\mathbf{k}^{\prime }\lambda }\left( \mathbf{q}\right) \left[ \mathbf{k}%
\cdot \mathbf{k}^{\prime }-\left( \mathbf{k}\cdot \mathbf{\hat{q}}\right)
\left( \mathbf{k}^{\prime }\cdot \mathbf{\hat{q}}\right) \right]  \notag \\
&&\times \sum_{ik_{m},ik_{m}^{\prime }}\mathcal{G}\left( \mathbf{k+q}%
,ik_{m}\right) \mathcal{G}\left( \mathbf{k},i\omega _{n}+ik_{m}\right) 
\mathcal{G}\left( \mathbf{k}^{\prime }\mathbf{+q},ik_{m}^{\prime }\right) 
\mathcal{G}\left( \mathbf{k}^{\prime },i\omega _{n}+ik_{m}^{\prime }\right) ,
\end{eqnarray}%
Summation over $ik_{m}$ and $ik_{m}^{\prime }$ leads to%
\begin{eqnarray}
\Pi ^{\left( 1\right) }\left( \mathbf{q},i\omega _{n}\right) &=&\frac{e^{2}}{%
m^{2}}D^{T}\left( \mathbf{q},i\omega _{n}\right) \sum_{\mathbf{kk}^{\prime
}}M_{\mathbf{k}\lambda }\left( \mathbf{q}\right) M_{\mathbf{k}^{\prime
}\lambda }\left( \mathbf{q}\right) \left[ \mathbf{k}\cdot \mathbf{k}^{\prime
}-\left( \mathbf{k}\cdot \mathbf{\hat{q}}\right) \left( \mathbf{k}^{\prime
}\cdot \mathbf{\hat{q}}\right) \right]  \notag \\
&&\times \int_{-\infty }^{\infty }\frac{d\epsilon }{\pi }n_{F}\left(
\epsilon \right) \left[ \mathcal{G}\left( \mathbf{k},\epsilon +i\omega
_{n}\right) A\left( \mathbf{k+q},\epsilon \right) +\mathcal{G}\left( \mathbf{%
k+q},\epsilon -i\omega _{n}\right) A\left( \mathbf{k},\epsilon \right) %
\right]  \notag \\
&&\times \int_{-\infty }^{\infty }\frac{d\epsilon ^{\prime }}{\pi }%
n_{F}\left( \epsilon ^{\prime }\right) \left[ \mathcal{G}\left( \mathbf{k}%
^{\prime },\epsilon ^{\prime }+i\omega _{n}\right) A\left( \mathbf{k}%
^{\prime }\mathbf{+q},\epsilon ^{\prime }\right) +\mathcal{G}\left( \mathbf{k%
}^{\prime }\mathbf{+q},\epsilon ^{\prime }-i\omega _{n}\right) A\left( 
\mathbf{k}^{\prime },\epsilon ^{\prime }\right) \right] ,
\end{eqnarray}%
where the spectral function $A\left( \mathbf{k},\epsilon \right) $ is given
by%
\begin{equation}
\mathcal{G}_{\text{ret}}\left( \mathbf{k},\epsilon \right) -\mathcal{G}_{%
\text{adv}}\left( \mathbf{k},\epsilon \right) =2i\text{Im}\mathcal{G}_{\text{%
ret}}\left( \mathbf{k},\epsilon \right) =-iA\left( \mathbf{k},\epsilon
\right) .
\end{equation}%
Analytically continue $i\omega _{n}\rightarrow \omega +i\eta $, we have%
\begin{eqnarray*}
\Pi _{\text{ret}}^{\left( 1\right) }\left( \mathbf{q},\omega \right) &=&%
\frac{e^{2}}{m^{2}}D^{T}\left( \mathbf{q},\omega \right) \sum_{\mathbf{kk}%
^{\prime }}M_{\mathbf{k}\lambda }\left( \mathbf{q}\right) M_{\mathbf{k}%
^{\prime }\lambda }\left( \mathbf{q}\right) \left[ \mathbf{k}\cdot \mathbf{k}%
^{\prime }-\left( \mathbf{k}\cdot \mathbf{\hat{q}}\right) \left( \mathbf{k}%
^{\prime }\cdot \mathbf{\hat{q}}\right) \right] \\
&&\times \int_{-\infty }^{\infty }\frac{d\epsilon }{\pi }n_{F}\left(
\epsilon \right) \left[ \mathcal{G}_{\text{ret}}\left( \mathbf{k},\epsilon
+\omega \right) A\left( \mathbf{k+q},\epsilon \right) +\mathcal{G}_{\text{adv%
}}\left( \mathbf{k+q},\epsilon -\omega \right) A\left( \mathbf{k},\epsilon
\right) \right] \\
&&\times \int_{-\infty }^{\infty }\frac{d\epsilon ^{\prime }}{\pi }%
n_{F}\left( \epsilon ^{\prime }\right) \left[ \mathcal{G}_{\text{ret}}\left( 
\mathbf{k}^{\prime },\epsilon ^{\prime }+\omega \right) A\left( \mathbf{k}%
^{\prime }\mathbf{+q},\epsilon ^{\prime }\right) +\mathcal{G}_{\text{adv}%
}\left( \mathbf{k}^{\prime }\mathbf{+q},\epsilon ^{\prime }-\omega \right)
A\left( \mathbf{k}^{\prime },\epsilon ^{\prime }\right) \right] .
\end{eqnarray*}%
After integrating over $\mathbf{k}$ and $\mathbf{k}^{\prime }$, $\mathcal{G}%
_{\text{ret}}\mathcal{G}_{\text{ret}}$ and $\mathcal{G}_{\text{adv}}\mathcal{%
G}_{\text{adv}}$ terms vanish, we obtain%
\begin{eqnarray}
\Pi _{\text{ret}}^{\left( 1\right) }\left( \mathbf{q},\omega \right) &=&-%
\frac{e^{2}}{m^{2}}D_{\text{ret}}^{T}\left( \mathbf{q},\omega \right) \sum_{%
\mathbf{kk}^{\prime }}M_{\mathbf{k}\lambda }\left( \mathbf{q}\right) M_{%
\mathbf{k}^{\prime }\lambda }\left( \mathbf{q}\right) \left[ \mathbf{k}\cdot 
\mathbf{k}^{\prime }-\left( \mathbf{k}\cdot \mathbf{\hat{q}}\right) \left( 
\mathbf{k}^{\prime }\cdot \mathbf{\hat{q}}\right) \right]  \notag \\
&&\times \int_{-\infty }^{\infty }\frac{d\epsilon }{\pi }n_{F}\left(
\epsilon \right) \left[ -\mathcal{G}_{\text{ret}}\left( \mathbf{k},\epsilon
+\omega \right) \mathcal{G}_{\text{adv}}\left( \mathbf{k+q},\epsilon \right)
+\mathcal{G}_{\text{adv}}\left( \mathbf{k+q},\epsilon -\omega \right) 
\mathcal{G}_{\text{ret}}\left( \mathbf{k},\epsilon \right) \right]  \notag \\
&&\times \int_{-\infty }^{\infty }\frac{d\epsilon ^{\prime }}{\pi }%
n_{F}\left( \epsilon ^{\prime }\right) \left[ -\mathcal{G}_{\text{ret}%
}\left( \mathbf{k}^{\prime },\epsilon ^{\prime }+\omega \right) \mathcal{G}_{%
\text{adv}}\left( \mathbf{k}^{\prime }\mathbf{+q},\epsilon ^{\prime }\right)
+\mathcal{G}_{\text{adv}}\left( \mathbf{k}^{\prime }\mathbf{+q},\epsilon
^{\prime }-\omega \right) \mathcal{G}_{\text{ret}}\left( \mathbf{k}^{\prime
},\epsilon ^{\prime }\right) \right]  \notag \\
&=&-\frac{e^{2}\omega ^{2}}{m^{2}}D_{\text{ret}}^{T}\left( \mathbf{q},\omega
\right) \sum_{\mathbf{kk}^{\prime }}M_{\mathbf{k}\lambda }\left( \mathbf{q}%
\right) M_{\mathbf{k}^{\prime }\lambda }\left( \mathbf{q}\right) \left[ 
\mathbf{k}\cdot \mathbf{k}^{\prime }-\left( \mathbf{k}\cdot \mathbf{\hat{q}}%
\right) \left( \mathbf{k}^{\prime }\cdot \mathbf{\hat{q}}\right) \right] 
\notag \\
&&\times \int_{-\infty }^{\infty }\frac{d\epsilon }{\pi }\left[ -\frac{%
\partial n_{F}\left( \epsilon \right) }{\partial \epsilon }\right] \mathcal{G%
}_{\text{ret}}\left( \mathbf{k},\epsilon +\omega \right) \mathcal{G}_{\text{%
adv}}\left( \mathbf{k+q},\epsilon \right)  \notag \\
&&\times \int_{-\infty }^{\infty }\frac{d\epsilon ^{\prime }}{\pi }\left[ -%
\frac{\partial n_{F}\left( \epsilon ^{\prime }\right) }{\partial \epsilon
^{\prime }}\right] \mathcal{G}_{\text{ret}}\left( \mathbf{k}^{\prime
},\epsilon ^{\prime }+\omega \right) \mathcal{G}_{\text{adv}}\left( \mathbf{k%
}^{\prime }\mathbf{+q},\epsilon ^{\prime }\right) .
\end{eqnarray}%
Therefore%
\begin{eqnarray}
\Pi _{\text{ret}}^{\left( 1\right) }\left( \mathbf{q},\omega \right) &=&-%
\frac{e^{2}k_{F}^{2}\omega }{2m^{2}}D_{\text{ret}}^{T}\left( \mathbf{q}%
,\omega \right) \frac{\omega ^{2}N\left( 0\right) k_{F}^{3}}{4\rho
_{ion}mv_{s}}  \notag \\
&&\times \frac{N\left( 0\right) }{qv_{F}}\left[ s_{1}\left( ql\right)
-s_{3}\left( ql\right) \right] ^{2}  \notag \\
&=&iF\left( \mathbf{q},\omega \right) \text{Im}\Pi _{\text{ret}}^{\left(
0\right) }\left( \mathbf{q},\omega \right) ,
\end{eqnarray}%
where%
\begin{equation}
F\left( \mathbf{q},\omega \right) =-\frac{e^{2}k_{F}^{2}\omega }{2m^{2}}%
\frac{N\left( 0\right) }{qv_{F}}D_{\text{ret}}^{T}\left( \mathbf{q},\omega
\right) ql\left[ s_{1}\left( ql\right) -s_{3}\left( ql\right) \right] .
\end{equation}

For EM field the propagator $D_{\text{ret}}^{T}$ is given by $D_{\text{ret}%
}^{\text{EM}}\left( \mathbf{q},\omega \right) $%
\begin{equation}
D_{\text{ret}}^{\text{EM}}=\frac{1}{i\omega \sigma _{\perp }\left( q,\omega
\right) +\omega ^{2}-c^{2}q^{2}}.
\end{equation}%
Following Pippard's notation, we write $\sigma _{\perp }\left( q,\omega
\right) =g\sigma _{0}$, where $\sigma _{0}=e^{2}n\tau /m$ is the DC
conductivity. To determine the factor $g$, we shall calculate $\sigma
_{\perp }\left( q,\omega \right) $ using the relation,%
\begin{equation}
\sigma _{\perp }\left( q,\omega \right) =-\frac{1}{\omega }\text{Im}\Pi
_{j\bot }\left( \mathbf{q},\omega \right) ,
\end{equation}%
where $\Pi _{j\bot }\left( \mathbf{q},i\omega _{n}\right) =\int_{0}^{\beta
}d\tau e^{i\omega _{n}\tau }\Pi _{j\bot }\left( \mathbf{q},\tau \right) $
and $\Pi _{j\bot }\left( \mathbf{q},\tau \right) =-\left\langle T_{\tau }%
\mathbf{j}_{\perp }\left( \mathbf{q},\tau \right) \cdot \mathbf{j}_{\perp
}\left( -\mathbf{q},0\right) \right\rangle $ the retarded correlation
function. To be simple, we shall further count $\Pi _{\bot }\left( \mathbf{q}%
,\omega \right) $ as the single bubble contribution. So that 
\begin{equation}
\Pi _{j\bot }\left( \mathbf{q},i\omega _{n}\right) =\frac{2e^{2}}{\beta m^{2}%
}\sum_{\mathbf{k}}\frac{1}{2}\left[ \mathbf{k-}\left( \mathbf{k\cdot \hat{q}}%
\right) \mathbf{\hat{q}}\right] ^{2}\sum_{ik_{m}}\mathcal{G}\left( \mathbf{%
k+q},ik_{m}\right) \mathcal{G}\left( \mathbf{k},i\omega _{n}+ik_{m}\right) .
\end{equation}%
Taking analytical continuation and then we have the imaginary part of $\Pi
_{j\bot }\left( \mathbf{q},\omega \right) $%
\begin{eqnarray}
\text{Im}\Pi _{j\bot }\left( \mathbf{q},\omega \right) &=&-\frac{\omega e^{2}%
}{m^{2}}\text{Re}\sum_{\mathbf{k}}[k^{2}\mathbf{-}\left( \mathbf{k\cdot \hat{%
q}}\right) ^{2}]\int_{-\infty }^{\infty }\frac{d\epsilon }{2\pi }\left[ -%
\frac{\partial n_{F}\left( \epsilon \right) }{\partial \epsilon }\right] 
\mathcal{G}_{\text{ret}}\left( \mathbf{k},\epsilon +\omega \right) \mathcal{G%
}_{\text{adv}}\left( \mathbf{k+q},\epsilon \right)  \notag \\
&=&\frac{\omega e^{2}k_{F}^{2}}{2\pi m^{2}}\frac{N\left( 0\right) \pi }{%
qv_{F}}\text{Re}\left[ s_{0}\left( a\right) -s_{2}\left( a\right) \right] .
\end{eqnarray}
Then we obtain%
\begin{eqnarray}
\sigma _{\perp }\left( q,\omega \right) &=&-\frac{e^{2}k_{F}^{2}}{2m^{2}}%
\frac{N\left( 0\right) }{qv_{F}}\text{Re}\left[ s_{0}\left( a\right)
-s_{2}\left( a\right) \right]  \notag \\
&=&\frac{3\sigma _{0}}{2ql}\left[ s_{2}\left( ql\right) -s_{0}\left(
ql\right) \right] ,
\end{eqnarray}%
where $\sigma _{0}=\frac{N\left( 0\right) e^{2}k_{F}^{2}\tau }{3m^{2}}$ is
used. So that the factor $g$ reads%
\begin{equation}
g=\frac{3}{2ql}\left[ s_{2}\left( ql\right) -s_{0}\left( ql\right) \right] .
\end{equation}%
We find that $g\rightarrow 1-\frac{2\left( ql\right) ^{2}}{15}$ when $ql\ll
1 $ and $g\rightarrow \frac{3\pi }{4ql}$ when $ql\gg 1$. We are interested
in the case when $c^{2}q^{2}\ll \omega \sigma _{\perp }\left( q,\omega
\right) $, say, $q\ll g^{-1/2}k_{0}$, where $k_{0}^{-1}$ is the classical
skin depth. This holds under the condition $q\ll k_{0}$ if $ql\ll 1$ and $%
q^{2}\ll k_{0}^{2}/\left( ql\right) $ if $ql\gg 1$. In this case,%
\begin{equation}
D_{\text{ret}}^{\text{EM}}=\frac{1}{i\omega \sigma _{\perp }\left( q,\omega
\right) },
\end{equation}%
and%
\begin{eqnarray}
F\left( \mathbf{q},\omega \right) &=&-\frac{e^{2}k_{F}^{2}\omega }{2m^{2}}%
\frac{N\left( 0\right) }{qv_{F}}\frac{ql\left[ s_{1}\left( ql\right)
-s_{3}\left( ql\right) \right] }{i\omega \sigma _{\perp }\left( q,\omega
\right) }  \notag \\
&=&\frac{3i}{2g}\left[ s_{1}\left( ql\right) -s_{3}\left( ql\right) \right] .
\end{eqnarray}%
Using the relation (\ref{itsn}), we have%
\begin{eqnarray*}
s_{1}\left( a\right) &=&-i-\frac{is_{0}\left( a\right) }{a}, \\
s_{3}\left( a\right) &=&-\frac{i}{3}-\frac{is_{2}\left( a\right) }{a}.
\end{eqnarray*}%
So that%
\begin{eqnarray}
s_{1}\left( ql\right) -s_{3}\left( ql\right) &=&-\frac{2i}{3}-i\frac{%
s_{0}\left( ql\right) -s_{2}\left( ql\right) }{ql}+O\left( \frac{v_{s}}{v_{F}%
}\right)  \notag \\
&=&-\frac{2i}{3}\left( 1-g\right) +O\left( \frac{v_{s}}{v_{F}}\right) ,
\end{eqnarray}%
and%
\begin{equation}
F=\frac{1-g}{g}+O\left( \frac{v_{s}}{v_{F}}\right) .
\end{equation}%
So that we obtain 
\begin{equation}
\Pi _{\text{ret}}^{\left( 1\right) }\left( \mathbf{q},\omega \right) =\frac{%
1-g}{g}\text{Im}\Pi _{\text{ret}}^{\left( 0\right) }\left( \mathbf{q},\omega
\right) +O\left( \frac{v_{s}}{v_{F}}\right) .
\end{equation}%
Thus the ultrasound attenuation coeffecient reads%
\begin{equation}
\alpha =-\frac{2}{v_{s}}\text{Im}(\Pi _{\text{ret}}^{(0)}+\Pi _{\text{ret}%
}^{(1)})=\frac{nm}{\rho _{ion}v_{s}\tau }\frac{1-g}{g}.
\end{equation}

\section{Transverse sound attenuation in a spin liquid}

For a spin liquid, the spinons and gauge fields are treated in 2D. The gauge
field propagator $D^{T} $ is replaced by%
\begin{equation}
D_{\text{ret}}^{T}=\frac{1}{i\omega \tilde{\sigma}_{\perp }\left( q,\omega
\right) -\chi q^{2}},
\end{equation}%
where $\tilde{\sigma}_{\perp }=\tilde{g}\tilde{\sigma}_{0}$, $\tilde{\sigma}%
_{0}=n\tau /m$, $n$ is the spinon density and $\chi =1/(24\pi m)$ is the
Landau diagramnetism. Note that the coupling constant to the gauge field has
been set to unity instead of $e$.

Similary, we have%
\begin{equation}
\text{Im}\tilde{\Pi}_{j\perp }\left( \mathbf{q},\omega \right) =\frac{\omega
e^{2}k_{F}^{2}}{\pi m^{2}}\frac{N\left( 0\right) \pi }{qv_{F}}\text{Re}\left[
t_{0}\left( ql\right) -t_{2}\left( ql\right) \right] ,
\end{equation}%
and%
\begin{eqnarray}
\tilde{\sigma}_{\perp }\left( q,\omega \right) &=&-\frac{1}{\omega }\text{Im}%
\tilde{\Pi}_{j\bot }\left( \mathbf{q},\omega \right)  \notag \\
&=&-\frac{e^{2}k_{F}^{2}}{m^{2}}\frac{N\left( 0\right) }{qv_{F}}\text{Re}%
\left[ t_{0}\left( ql\right) -t_{2}\left( ql\right) \right]  \notag \\
&=&-\frac{2\tilde{\sigma}_{0}}{ql}\text{Re}\left[ t_{0}\left( ql\right)
-t_{2}\left( ql\right) \right] .
\end{eqnarray}%
So that the factor $\tilde{g}$ reads%
\begin{equation}
\tilde{g}=-\frac{2}{ql}\text{Re}\left[ t_{0}\left( ql\right) -t_{2}\left(
ql\right) \right] .
\end{equation}%
We also find that $\tilde{g}\rightarrow 1-\frac{\left( ql\right) ^{2}}{4}$
when $ql\ll 1$ and $\tilde{g}\rightarrow \frac{2}{ql}$ when $ql\gg 1$.

The imaginary part of $\tilde{\Pi}_{\text{ret}}^{(0)}$ is given by%
\begin{equation}
\text{Im}\tilde{\Pi}_{\text{ret}}^{\left( 0\right) }\left( \mathbf{q},\omega
\right) =\frac{\omega N\left( 0\right) k_{F}^{3}}{2\rho _{ion}mv_{s}}\frac{%
t_{1}\left( ql\right) -t_{3}\left( ql\right) }{iql},
\end{equation}%
and $\tilde{\Pi}_{\text{ret}}^{(1)}$ is given by%
\begin{eqnarray}
\tilde{\Pi}_{\text{ret}}^{\left( 1\right) }\left( \mathbf{q},\omega \right)
&=&(1+\tilde{F})\text{Im}\tilde{\Pi}_{\text{ret}}^{\left( 0\right) }\left( 
\mathbf{q},\omega \right) , \\
\tilde{F} &=&\frac{2i}{\tilde{g}}\left[ t_{1}\left( ql\right) -t_{3}\left(
ql\right) \right] .
\end{eqnarray}%
So that we still can write $\tilde{\Pi}_{\text{ret}}^{\left( 1\right) }$ as%
\begin{equation}
\tilde{\Pi}_{\text{ret}}^{\left( 1\right) }=\frac{1-\tilde{g}}{\tilde{g}}%
\text{Im}\tilde{\Pi}_{\text{ret}}^{\left( 0\right) }+O(\frac{v_{s}}{v_{F}})%
\text{.}
\end{equation}

\subsection{Onset of superconductivity}

For transeverse ultrasound, a rapid fall of the ratio $\alpha /\alpha _{N}$
below $T_{c}$ due to $\Pi ^{\left( 1\right) }$ has been discussed in the
paper in the clean limit $ql\gg 1$. We shall focus on $\Pi ^{\left( 0\right)
}$ term, which gives rise to the ratio $\alpha ^{(0)}/\alpha _{N}^{(0)}$ and 
$\alpha ^{(0)}$ is given by%
\begin{equation}
\alpha ^{(0)}=-\frac{2}{v_{s}}\text{Im}\Pi _{\text{ret}}^{(0)}.
\end{equation}%
$\Pi ^{\left( 0\right) }$ will decrease below $T_{c}$ because of the opening
of the energy gap.

During the process of the attenuation of a sound wave in such a fermionic
system, the sound wave passes its energy to fermions by fermion phonon
coupling and fermions dissipate this extra energy due to disorder to
complete the attenuation. Therefore the fermion life time $\tau $ should be
finite to relax the sound wave. On the other hand $v_{s}\ll v_{F}$, it is
different from many other relaxation mechanisms. So that despite $ql\gtrsim 1
$, the physically relevant situation is reached by $\omega \tau \ll 1$. It
means we can not take $\tau \rightarrow \infty $ at first even in the clean
limit because finite $\tau $ is still seen for $\omega \tau \ll 1$.
Otherwise, we would obtain a $\Pi ^{\left( 0\right) }$ which does not depend
on $ql$ and conflicts with Eq.(\ref{PiT02}). It is interesting that when
screening is included by $\Pi ^{\left( 0\right) }+\Pi ^{\left( 1\right) }$,
the result does not depend on $\tau $ and appears to agree with the infinite 
$\tau $ calculation (see Eq.(\ref{ImPi0})). It is only when screening is suppressed by
Meissner effect that we can see this distinction. Below we shall discuss in
the limit $\omega \tau \ll 1\ll \Delta \tau $, which is most interesting for
us. Note that our results differ from those in early literatures
(e.g., see J. R. Cullen and R. A. Ferrell, Phys. Rev. \textbf{146}, 282 (1966).), 
since they worked in the clean limit by setting $\tau
\rightarrow \infty $ which is not valid for $\omega \tau \ll 1$.

To study the pairing state with finite $\tau $, we shall adopt the following
impurity avaraged Green's function,%
\begin{eqnarray*}
\mathcal{G}\left( \mathbf{k},i\omega _{n}\right) &=&\frac{i\tilde{\omega}%
_{n}+\xi _{\mathbf{k}}}{\left( i\tilde{\omega}_{n}\right) ^{2}-\xi _{\mathbf{%
k}}^{2}-\left[ \tilde{\Delta}_{\mathbf{k}}\left( \omega _{n}\right) \right]
^{2}}, \\
\mathcal{F}\left( \mathbf{k},i\omega _{n}\right) &=&-\frac{\tilde{\Delta}_{%
\mathbf{k}}\left( \omega _{n}\right) }{\left( i\tilde{\omega}_{n}\right)
^{2}-\xi _{\mathbf{k}}^{2}-\left[ \tilde{\Delta}_{\mathbf{k}}\left( \omega
_{n}\right) \right] ^{2}}, \\
\tilde{\omega}_{n} &=&\omega _{n}(1+\frac{1}{2\tau \sqrt{\omega
_{n}^{2}+\Delta _{\mathbf{k}}^{2}}}), \\
\tilde{\Delta}_{\mathbf{k}}\left( \omega _{n}\right) &=&\Delta _{\mathbf{k}%
}(1+\frac{1}{2\tau \sqrt{\omega _{n}^{2}+\Delta _{\mathbf{k}}^{2}}}),
\end{eqnarray*}%
for superconducting states and replace the terms with $\mathcal{GG}$ by $%
\mathcal{GG}-\mathcal{FF}$. The results should coincide to Eq.(\ref{PiT02})
when $\Delta _{\mathbf{k}}\rightarrow 0$. We do frequency summation firstly.%
\begin{eqnarray}
S &=&\frac{1}{\beta }\sum_{ik_{m}}\mathcal{G}\left( \mathbf{k}%
,ik_{m}+i\omega _{n}\right) \mathcal{G}\left( \mathbf{k+q},ik_{m}\right) -%
\mathcal{F}\left( \mathbf{k},ik_{m}+i\omega _{n}\right) \mathcal{F}\left( 
\mathbf{k+q},ik_{m}\right)  \notag \\
&=&-\int_{C}\frac{dz}{2\pi i}n_{F}\left( z\right) \left[ \mathcal{G}\left( 
\mathbf{k},\epsilon +i\omega _{n}\right) \mathcal{G}\left( \mathbf{k+q}%
,z\right) -\mathcal{F}\left( \mathbf{k},\epsilon +i\omega _{n}\right) 
\mathcal{F}\left( \mathbf{k+q},\epsilon \right) \right]  \notag \\
&=&-\int_{-\infty }^{\infty }\frac{d\epsilon }{2\pi i}n_{F}\left( \epsilon
\right) \left\{ \mathcal{G}\left( \mathbf{k},\epsilon +i\omega _{n}\right) %
\left[ \mathcal{G}\left( \mathbf{k+q},\epsilon +i\eta \right) -\mathcal{G}%
\left( \mathbf{k+q},\epsilon -i\eta \right) \right] \right.  \notag \\
&&+\mathcal{G}\left( \mathbf{k+q},\epsilon -i\omega _{n}\right) \left[ 
\mathcal{G}\left( \mathbf{k},\epsilon +i\eta \right) -\mathcal{G}\left( 
\mathbf{k},\epsilon -i\eta \right) \right]  \notag \\
&&-\mathcal{F}\left( \mathbf{k},\epsilon +i\omega _{n}\right) \left[ 
\mathcal{F}\left( \mathbf{k+q},\epsilon +i\eta \right) -\mathcal{F}\left( 
\mathbf{k+q},\epsilon -i\eta \right) \right]  \notag \\
&&\left. -\mathcal{F}\left( \mathbf{k+q},\epsilon -i\omega _{n}\right) \left[
\mathcal{F}\left( \mathbf{k},\epsilon +i\eta \right) -\mathcal{F}\left( 
\mathbf{k},\epsilon -i\eta \right) \right] \right\} .
\end{eqnarray}%
The interval $\left( -\Delta ,\Delta \right) $, for which the Green's
functions are continuous across the real $\epsilon $ axis, can be excluded
from the $\epsilon $ integral. Taking $i\omega _{n}\rightarrow \omega +i\eta 
$, we have%
\begin{eqnarray}
S &=&-\int_{-\infty }^{\infty }\frac{d\epsilon }{2\pi i}n_{F}\left( \epsilon
\right) \left\{ \mathcal{G}_{\text{ret}}\left( \mathbf{k},\epsilon +\omega
\right) \left[ \mathcal{G}_{\text{ret}}\left( \mathbf{k+q},\epsilon \right) -%
\mathcal{G}_{\text{adv}}\left( \mathbf{k+q},\epsilon \right) \right] \right.
\notag \\
&&+\mathcal{G}_{\text{adv}}\left( \mathbf{k+q},\epsilon -\omega \right) %
\left[ \mathcal{G}_{\text{ret}}\left( \mathbf{k},\epsilon \right) -\mathcal{G%
}_{\text{adv}}\left( \mathbf{k},\epsilon \right) \right]  \notag \\
&&-\mathcal{F}_{\text{ret}}\left( \mathbf{k},\epsilon +\omega \right) \left[ 
\mathcal{F}_{\text{ret}}\left( \mathbf{k+q},\epsilon \right) -\mathcal{F}_{%
\text{adv}}\left( \mathbf{k+q},\epsilon \right) \right]  \notag \\
&&\left. -\mathcal{F}_{\text{adv}}\left( \mathbf{k+q},\epsilon -\omega
\right) \left[ \mathcal{F}_{\text{ret}}\left( \mathbf{k},\epsilon \right) -%
\mathcal{F}_{\text{adv}}\left( \mathbf{k},\epsilon \right) \right] \right\} ,
\end{eqnarray}%
where%
\begin{eqnarray*}
\mathcal{G}_{\text{ret(adv)}}\left( \mathbf{k},\epsilon \right) &=&\frac{%
\epsilon _{\mathbf{k}\pm }^{\prime }+\xi _{\mathbf{k}}}{\epsilon _{\mathbf{k}%
\pm }^{\prime 2}-\xi _{\mathbf{k}}^{2}-\Delta _{\mathbf{k}\pm }^{\prime 2}},
\\
\mathcal{F}_{\text{ret(adv)}}\left( \mathbf{k},\epsilon \right) &=&-\frac{%
\Delta _{\mathbf{k}\pm }^{\prime }}{\epsilon _{\mathbf{k}\pm }^{\prime
2}-\xi _{\mathbf{k}}^{2}-\Delta _{\mathbf{k}\pm }^{\prime 2}}, \\
\mathcal{G}_{\text{ret(adv)}}\left( \mathbf{k},\epsilon +\omega \right) &=&%
\frac{\epsilon _{\mathbf{k}\pm }^{\prime \prime }+\xi _{\mathbf{k}}}{%
\epsilon _{\mathbf{k}\pm }^{\prime \prime 2}-\xi _{\mathbf{k}}^{2}-\Delta _{%
\mathbf{k}\pm }^{\prime \prime 2}}, \\
\mathcal{F}_{\text{ret(adv)}}\left( \mathbf{k},\epsilon +\omega \right) &=&-%
\frac{\Delta _{\mathbf{k}\pm }^{\prime \prime }}{\epsilon _{\mathbf{k}\pm
}^{\prime \prime 2}-\xi _{\mathbf{k}}^{2}-\Delta _{\mathbf{k}\pm }^{\prime
\prime 2}}
\end{eqnarray*}%
with%
\begin{eqnarray*}
\epsilon _{\mathbf{k}\pm }^{\prime } &=&\epsilon \left( 1\pm \frac{i}{2\tau 
\sqrt{\epsilon ^{2}-\Delta _{\mathbf{k}}^{2}}}\right) , \\
\Delta _{\mathbf{k}\pm }^{\prime } &=&\Delta _{\mathbf{k}}\left( 1\pm \frac{i%
}{2\tau \sqrt{\epsilon ^{2}-\Delta _{\mathbf{k}}^{2}}}\right) , \\
\epsilon _{\mathbf{k}\pm }^{\prime \prime } &=&\left( \epsilon +\omega
\right) \left( 1\pm \frac{i}{2\tau \sqrt{\left( \epsilon +\omega \right)
^{2}-\Delta _{\mathbf{k}}^{2}}}\right) , \\
\Delta _{\mathbf{k}\pm }^{\prime \prime } &=&\Delta _{\mathbf{k}}\left( 1\pm 
\frac{i}{2\tau \sqrt{\left( \epsilon +\omega \right) ^{2}-\Delta _{\mathbf{k}%
}^{2}}}\right) .
\end{eqnarray*}%
Then we perform integation over $\mathbf{k}$ in $\Pi ^{(0)}$ which can
reduced to an integral over $\nu =\cos \theta $ and $\xi _{\mathbf{k}}$. To
do this, we write $\mathcal{G}_{\text{ret(adv)}}\left( \mathbf{k},\epsilon
\right) $ and $\mathcal{F}_{\text{ret(adv)}}\left( \mathbf{k},\epsilon
\right) $ in terms of the poles of $\xi _{\mathbf{k}}$, 
\begin{eqnarray*}
\mathcal{G}_{\text{ret(adv)}}\left( \mathbf{k},\epsilon \right) &=&\frac{%
\epsilon _{\mathbf{k}\pm }^{\prime }+\xi _{\mathbf{k}}}{\epsilon _{\mathbf{k}%
\pm }^{\prime 2}-\xi _{\mathbf{k}}^{2}-\Delta _{\mathbf{k}\pm }^{\prime 2}}=%
\frac{\epsilon _{\mathbf{k}\pm }^{\prime }+\xi _{\mathbf{k}}}{2\zeta _{%
\mathbf{k\pm }}^{\prime }}\left[ \frac{1}{\zeta _{\mathbf{k\pm }}^{\prime
}-\xi _{\mathbf{k}}}+\frac{1}{\zeta _{\mathbf{k\pm }}^{\prime }+\xi _{%
\mathbf{k}}}\right] , \\
\mathcal{F}_{\text{ret(adv)}}\left( \mathbf{k},\epsilon \right) &=&-\frac{%
\Delta _{\mathbf{k}\pm }^{\prime }}{\epsilon _{\mathbf{k}\pm }^{\prime
2}-\xi _{\mathbf{k}}^{2}-\Delta _{\mathbf{k}\pm }^{\prime 2}}=-\frac{\Delta
_{\mathbf{k}\pm }^{\prime }}{2\zeta _{\mathbf{k\pm }}}\left[ \frac{1}{\zeta
_{\mathbf{k\pm }}^{\prime }-\xi _{\mathbf{k}}}+\frac{1}{\zeta _{\mathbf{k\pm 
}}^{\prime }+\xi _{\mathbf{k}}}\right] , \\
\mathcal{G}_{\text{ret(adv)}}\left( \mathbf{k},\epsilon +\omega \right) &=&%
\frac{\epsilon _{\mathbf{k}\pm }^{\prime \prime }+\xi _{\mathbf{k}}}{%
\epsilon _{\mathbf{k}\pm }^{\prime \prime 2}-\xi _{\mathbf{k}}^{2}-\Delta _{%
\mathbf{k}\pm }^{\prime \prime 2}}=\frac{\epsilon _{\mathbf{k}\pm }^{\prime
\prime }+\xi _{\mathbf{k}}}{2\zeta _{\mathbf{k\pm }}^{\prime \prime }}\left[ 
\frac{1}{\zeta _{\mathbf{k\pm }}^{\prime \prime }-\xi _{\mathbf{k}}}+\frac{1%
}{\zeta _{\mathbf{k\pm }}^{\prime \prime }+\xi _{\mathbf{k}}}\right] , \\
\mathcal{F}_{\text{ret(adv)}}\left( \mathbf{k},\epsilon +\omega \right) &=&-%
\frac{\Delta _{\mathbf{k}\pm }^{\prime \prime }}{\epsilon _{\mathbf{k}\pm
}^{\prime \prime 2}-\xi _{\mathbf{k}}^{2}-\Delta _{\mathbf{k}\pm }^{\prime
\prime 2}}=-\frac{\Delta _{\mathbf{k}\pm }^{\prime \prime }}{2\zeta _{%
\mathbf{k\pm }}}\left[ \frac{1}{\zeta _{\mathbf{k\pm }}^{\prime \prime }-\xi
_{\mathbf{k}}}+\frac{1}{\zeta _{\mathbf{k\pm }}^{\prime \prime }+\xi _{%
\mathbf{k}}}\right] ,
\end{eqnarray*}%
where%
\begin{eqnarray*}
\zeta _{\mathbf{k\pm }}^{\prime } &=&\sqrt{\epsilon _{\mathbf{k}\pm
}^{\prime 2}-\Delta _{\mathbf{k}\pm }^{\prime 2}}=\zeta _{\mathbf{k}%
}^{\prime }\pm \frac{i}{2\tau }, \\
\zeta _{\mathbf{k\pm }}^{\prime \prime } &=&\sqrt{\epsilon _{\mathbf{k}\pm
}^{\prime \prime 2}-\Delta _{\mathbf{k}\pm }^{\prime \prime 2}}=\zeta _{%
\mathbf{k}}^{\prime \prime }\pm \frac{i}{2\tau }, \\
\zeta _{\mathbf{k}}^{\prime } &=&\sqrt{\epsilon ^{2}-\Delta _{\mathbf{k}}^{2}%
}, \\
\zeta _{\mathbf{k}}^{\prime \prime } &=&\sqrt{\left( \epsilon +\omega
\right) ^{2}-\Delta _{\mathbf{k}}^{2}}.
\end{eqnarray*}%
We find that the following relations will be useful in determing the
imaginary part of $\Pi ^{(0)}$,%
\begin{eqnarray*}
\frac{\epsilon _{\mathbf{k}\pm }^{\prime }}{\zeta _{\mathbf{k\pm }}^{\prime }%
} &=&\frac{\epsilon }{\sqrt{\epsilon ^{2}-\Delta _{\mathbf{k}}^{2}}}, \\
\frac{\Delta _{\mathbf{k}\pm }^{\prime }}{\zeta _{\mathbf{k\pm }}^{\prime }}
&=&\frac{\Delta _{\mathbf{k}}}{\sqrt{\epsilon ^{2}-\Delta _{\mathbf{k}}^{2}}}%
, \\
\frac{\epsilon _{\mathbf{k}\pm }^{\prime \prime }}{\zeta _{\mathbf{k\pm }%
}^{\prime \prime }} &=&\frac{\epsilon +\omega }{\sqrt{\epsilon ^{2}-\Delta _{%
\mathbf{k}}^{2}}}, \\
\frac{\Delta _{\mathbf{k}\pm }^{\prime \prime }}{\zeta _{\mathbf{k\pm }%
}^{\prime \prime }} &=&\frac{\Delta _{\mathbf{k}}}{\sqrt{\left( \epsilon
+\omega \right) ^{2}-\Delta _{\mathbf{k}}^{2}}}.
\end{eqnarray*}%
Integrating with respect to $\xi _{\mathbf{k}}$ leads to%
\begin{eqnarray*}
&&\int \frac{d\xi _{\mathbf{k}}}{\pi i}\mathcal{G}_{\text{adv}}\left( 
\mathbf{k+q},\epsilon \right) \mathcal{G}_{\text{ret}}\left( \mathbf{k}%
,\epsilon +\omega \right) \\
&=&\int \frac{d\xi _{\mathbf{k}}}{\pi i}\frac{\epsilon _{\mathbf{k+q}%
-}^{\prime }+\xi _{\mathbf{k}}+qv_{F}\nu }{2\zeta _{\mathbf{k+q-}}^{\prime }}%
\frac{\epsilon _{\mathbf{k}+}^{\prime \prime }+\xi _{\mathbf{k}}}{2\zeta _{%
\mathbf{k+}}^{\prime \prime }} \\
&&\times \left[ \frac{1}{\zeta _{\mathbf{k+q-}}^{\prime }-\xi _{\mathbf{k}%
}-qv_{F}\nu }+\frac{1}{\zeta _{\mathbf{k+q-}}^{\prime }+\xi _{\mathbf{k}%
}+qv_{F}\nu }\right] \left[ \frac{1}{\zeta _{\mathbf{k+}}^{\prime \prime
}-\xi _{\mathbf{k}}}+\frac{1}{\zeta _{\mathbf{k+}}^{\prime \prime }+\xi _{%
\mathbf{k}}}\right] \\
&=&\frac{1}{2\zeta _{\mathbf{k+q-}}^{\prime }\zeta _{\mathbf{k+}}^{\prime
\prime }}\frac{\left( \epsilon _{\mathbf{k+q}-}^{\prime }+\zeta _{\mathbf{%
k+q-}}^{\prime }\right) \left( \epsilon _{\mathbf{k}+}^{\prime \prime
}+\zeta _{\mathbf{k+}}^{\prime \prime }\right) }{\zeta _{\mathbf{k+}%
}^{\prime \prime }-\zeta _{\mathbf{k+q-}}^{\prime }+qv_{F}\nu }+\frac{1}{%
2\zeta _{\mathbf{k+q-}}^{\prime }\zeta _{\mathbf{k+}}^{\prime \prime }}\frac{%
-\left( \epsilon _{\mathbf{k+q}-}^{\prime }-\zeta _{\mathbf{k+q-}}^{\prime
}\right) \left( \epsilon _{\mathbf{k}+}^{\prime \prime }-\zeta _{\mathbf{k+}%
}^{\prime \prime }\right) }{\zeta _{\mathbf{k+}}^{\prime \prime }-\zeta _{%
\mathbf{k+q-}}^{\prime }-qv_{F}\nu } \\
&&+\frac{1}{2\zeta _{\mathbf{k+q-}}^{\prime }\zeta _{\mathbf{k+}}^{\prime
\prime }}\frac{-\left( \epsilon _{\mathbf{k+q}-}^{\prime }-\zeta _{\mathbf{%
k+q-}}^{\prime }\right) \left( \epsilon _{\mathbf{k}+}^{\prime \prime
}+\zeta _{\mathbf{k+}}^{\prime \prime }\right) }{\zeta _{\mathbf{k+}%
}^{\prime \prime }+\zeta _{\mathbf{k+q-}}^{\prime }+qv_{F}\nu }+\frac{1}{%
2\zeta _{\mathbf{k+q-}}^{\prime }\zeta _{\mathbf{k+}}^{\prime \prime }}\frac{%
\left( \epsilon _{\mathbf{k+q}-}^{\prime }+\zeta _{\mathbf{k+q-}}^{\prime
}\right) \left( \epsilon _{\mathbf{k}+}^{\prime \prime }-\zeta _{\mathbf{k+}%
}^{\prime \prime }\right) }{\zeta _{\mathbf{k+}}^{\prime \prime }+\zeta _{%
\mathbf{k+q-}}^{\prime }-qv_{F}\nu }
\end{eqnarray*}%
where%
\begin{equation*}
\int \frac{d\xi _{\mathbf{k}}}{\pi i}F\left( \xi _{\mathbf{k}}\right) \frac{1%
}{\zeta _{\mathbf{k\pm }}^{\prime }-\xi _{\mathbf{k}}}=\mp F\left( \zeta _{%
\mathbf{k\pm }}^{\prime }\right)
\end{equation*}%
etc. is used. Similarly, we have 
\begin{eqnarray*}
&&\int \frac{d\xi _{\mathbf{k}}}{\pi i}\mathcal{F}_{\text{adv}}\left( 
\mathbf{k+q},\epsilon \right) \mathcal{F}_{\text{ret}}\left( \mathbf{k}%
,\epsilon +\omega \right) \\
&=&\frac{1}{2\zeta _{\mathbf{k+q-}}^{\prime }\zeta _{\mathbf{k+}}^{\prime
\prime }}\frac{\Delta _{\mathbf{k+q}-}^{\prime }\Delta _{\mathbf{k}%
+}^{\prime \prime }}{\zeta _{\mathbf{k+}}^{\prime \prime }-\zeta _{\mathbf{%
k+q-}}^{\prime }+qv_{F}\nu }-\frac{1}{2\zeta _{\mathbf{k+q-}}^{\prime }\zeta
_{\mathbf{k+}}^{\prime \prime }}\frac{\Delta _{\mathbf{k+q}-}^{\prime
}\Delta _{\mathbf{k}+}^{\prime \prime }}{\zeta _{\mathbf{k+}}^{\prime \prime
}-\zeta _{\mathbf{k+q-}}^{\prime }-qv_{F}\nu } \\
&&-\frac{1}{2\zeta _{\mathbf{k+q-}}^{\prime }\zeta _{\mathbf{k+}}^{\prime
\prime }}\frac{\Delta _{\mathbf{k+q}-}^{\prime }\Delta _{\mathbf{k}%
+}^{\prime \prime }}{\zeta _{\mathbf{k+}}^{\prime \prime }+\zeta _{\mathbf{%
k+q-}}^{\prime }+qv_{F}\nu }+\frac{1}{2\zeta _{\mathbf{k+q-}}^{\prime }\zeta
_{\mathbf{k+}}^{\prime \prime }}\frac{\Delta _{\mathbf{k+q}-}^{\prime
}\Delta _{\mathbf{k}+}^{\prime \prime }}{\zeta _{\mathbf{k+}}^{\prime \prime
}+\zeta _{\mathbf{k+q-}}^{\prime }-qv_{F}\nu }.
\end{eqnarray*}%
So that%
\begin{eqnarray}
&&\int \frac{d\xi _{\mathbf{k}}}{\pi i}\left[ \mathcal{G}_{\text{adv}}\left( 
\mathbf{k+q},\epsilon \right) \mathcal{G}_{\text{ret}}\left( \mathbf{k}%
,\epsilon +\omega \right) -\mathcal{F}_{\text{adv}}\left( \mathbf{k+q}%
,\epsilon \right) \mathcal{F}_{\text{ret}}\left( \mathbf{k},\epsilon +\omega
\right) \right]  \notag \\
&=&\frac{1}{2\zeta _{\mathbf{k+q-}}^{\prime }\zeta _{\mathbf{k+}}^{\prime
\prime }}\frac{\left( \epsilon _{\mathbf{k+q}-}^{\prime }+\zeta _{\mathbf{%
k+q-}}^{\prime }\right) \left( \epsilon _{\mathbf{k}+}^{\prime \prime
}+\zeta _{\mathbf{k+}}^{\prime \prime }\right) -\Delta _{\mathbf{k+q}%
-}^{\prime }\Delta _{\mathbf{k}+}^{\prime \prime }}{\zeta _{\mathbf{k+}%
}^{\prime \prime }-\zeta _{\mathbf{k+q-}}^{\prime }+qv_{F}\nu }  \notag \\
&&-\frac{1}{2\zeta _{\mathbf{k+q-}}^{\prime }\zeta _{\mathbf{k+}}^{\prime
\prime }}\frac{\left( \epsilon _{\mathbf{k+q}-}^{\prime }-\zeta _{\mathbf{%
k+q-}}^{\prime }\right) \left( \epsilon _{\mathbf{k}+}^{\prime \prime
}-\zeta _{\mathbf{k+}}^{\prime \prime }\right) -\Delta _{\mathbf{k+q}%
-}^{\prime }\Delta _{\mathbf{k}+}^{\prime \prime }}{\zeta _{\mathbf{k+}%
}^{\prime \prime }-\zeta _{\mathbf{k+q-}}^{\prime }-qv_{F}\nu }  \notag \\
&&-\frac{1}{2\zeta _{\mathbf{k+q-}}^{\prime }\zeta _{\mathbf{k+}}^{\prime
\prime }}\frac{\left( \epsilon _{\mathbf{k+q}-}^{\prime }-\zeta _{\mathbf{%
k+q-}}^{\prime }\right) \left( \epsilon _{\mathbf{k}+}^{\prime \prime
}+\zeta _{\mathbf{k+}}^{\prime \prime }\right) -\Delta _{\mathbf{k+q}%
-}^{\prime }\Delta _{\mathbf{k}+}^{\prime \prime }}{\zeta _{\mathbf{k+}%
}^{\prime \prime }+\zeta _{\mathbf{k+q-}}^{\prime }+qv_{F}\nu }  \notag \\
&&+\frac{1}{2\zeta _{\mathbf{k+q-}}^{\prime }\zeta _{\mathbf{k+}}^{\prime
\prime }}\frac{\left( \epsilon _{\mathbf{k+q}-}^{\prime }+\zeta _{\mathbf{%
k+q-}}^{\prime }\right) \left( \epsilon _{\mathbf{k}+}^{\prime \prime
}-\zeta _{\mathbf{k+}}^{\prime \prime }\right) -\Delta _{\mathbf{k+q}%
-}^{\prime }\Delta _{\mathbf{k}+}^{\prime \prime }}{\zeta _{\mathbf{k+}%
}^{\prime \prime }+\zeta _{\mathbf{k+q-}}^{\prime }-qv_{F}\nu }.
\end{eqnarray}%
For the terms of $\mathcal{G}_{\text{ret}}\mathcal{G}_{\text{ret}}-\mathcal{F%
}_{\text{ret}}\mathcal{F}_{\text{ret}}$ and $\mathcal{G}_{\text{adv}}%
\mathcal{G}_{\text{adv}}-\mathcal{F}_{\text{adv}}\mathcal{F}_{\text{adv}}$,
we need only to replace $\epsilon _{-}^{\prime }$ and $\zeta _{\mathbf{-}%
}^{\prime }$ by $\epsilon _{+}^{\prime }$ and $\zeta _{\mathbf{+}}^{\prime }$
in the above, so on and so forth. Then we find that only the terms of $%
\mathcal{G}_{\text{adv}}\mathcal{G}_{\text{ret}}-\mathcal{F}_{\text{adv}}%
\mathcal{F}_{\text{ret}}$ contribute to the imaginary part of $\Pi _{\text{%
ret}}^{(0)}$. Considering $\left[ M_{\mathbf{k}\lambda }\left( \mathbf{q}%
\right) \right] ^{2}$ is an even function of $\nu $, we have%
\begin{eqnarray}
\text{Im}\Pi _{\text{ret}}^{(0)} &=&2\omega \pi N\left( 0\right) \text{Re}%
\int_{-1}^{1}d\nu \int_{\left\vert \Delta _{\mathbf{k}}\right\vert }^{\infty
}\frac{d\epsilon }{2\pi }\frac{[M_{\mathbf{k}\lambda }\left( \mathbf{q}%
\right) ]^{2}\left[ 1+\frac{\epsilon \left( \epsilon +\omega \right) -\Delta
_{\mathbf{k}}\Delta _{\mathbf{k+q}}}{\sqrt{\epsilon ^{2}-\Delta _{\mathbf{k+q%
}}^{2}}\sqrt{\left( \epsilon +\omega \right) ^{2}-\Delta _{\mathbf{k}}^{2}}}%
\right] }{\sqrt{\left( \epsilon +\omega \right) ^{2}-\Delta _{\mathbf{k}}^{2}%
}-\sqrt{\epsilon ^{2}-\Delta _{\mathbf{k+q}}^{2}}+i/\tau +qv_{F}\nu }\left[ -%
\frac{\partial n_{F}\left( \epsilon \right) }{\partial \epsilon }\right] 
\notag \\
&\simeq &2\omega \pi N\left( 0\right) \text{Re}\int_{-1}^{1}d\nu
\int_{\left\vert \Delta _{\mathbf{k}}\right\vert }^{\infty }\frac{d\epsilon 
}{2\pi }\left[ -\frac{\partial n_{F}\left( \epsilon \right) }{\partial
\epsilon }\right] \frac{[M_{\mathbf{k}\lambda }\left( \mathbf{q}\right) ]^{2}%
\left[ 1+\frac{\epsilon ^{2}-\Delta _{\mathbf{k}}\Delta _{\mathbf{k+q}}}{%
\sqrt{\epsilon ^{2}-\Delta _{\mathbf{k+q}}^{2}}\sqrt{\epsilon ^{2}-\Delta _{%
\mathbf{k}}^{2}}}\right] }{\sqrt{\epsilon ^{2}-\Delta _{\mathbf{k}}^{2}}-%
\sqrt{\epsilon ^{2}-\Delta _{\mathbf{k+q}}^{2}}+i/\tau +qv_{F}\nu }.
\end{eqnarray}%
For a $s$-wave pairing state, Im$\Pi _{\text{ret}}^{(0)}$ can be evaluated
explicitly,%
\begin{eqnarray}
\text{Im}\Pi _{\text{ret}}^{(0)} &\simeq &2\omega \pi N\left( 0\right) \text{%
Re}\int_{\Delta }^{\infty }\frac{d\epsilon }{2\pi }\left[ -\frac{\partial
n_{F}\left( \epsilon \right) }{\partial \epsilon }\right] \int_{-1}^{1}d\nu 
\frac{\lbrack M_{\mathbf{k}\lambda }\left( \mathbf{q}\right) ]^{2}}{%
qv_{F}\nu +i/\tau }  \notag \\
&=&2n_{F}\left( \Delta \right) \text{Im}\Pi _{N}^{(0)}.
\end{eqnarray}%
The ratio $\alpha ^{(0)}/\alpha _{N}^{(0)}$ is given by%
\begin{equation}
\frac{\alpha ^{(0)}}{\alpha _{N}^{(0)}}=2n_{F}\left( \Delta \right) ,
\end{equation}%
in the limit $\omega \tau \ll 1\ll \Delta \tau .$
\end{widetext}

%\setcounter{bibitem}[21]

%\begin{thebibliography}
%\bibitem{CullenFerrell} J. R. Cullen and R. A. Ferrell, Phys. Rev. \textbf{146}, 282 (1966).
%\end{thebibliography}
\end{document}